\documentstyle[12pt]{article}
\normalbaselines
\begin{document}
\begin{center}
{\bf Multiparticle production and quantum chromodynamics}

\bigskip

I.M. Dremin

\bigskip

{\it Lebedev Physical Institute, Moscow 119991, Russia}

\end{center}

\begin{abstract}
The theory of strong interactions, quantum chromodynamics (QCD), is quite
successful in the prediction and description of main features of multiparticle
production processes at high energies. The general perturbative QCD approach
to these processes (mainly to $e^{+}e^{-}$-annihilation) is briefly formulated 
and its problems are discussed. It is shown that the analytical calculations
at the parton level with the low-momentum cut-off reproduce experimental data
on the hadronic final state in multiparticle production processes at high
energies surprisingly accurately even though the perturbative
expansion parameter is not very small. Moreover, it is important that
the perturbative QCD has been able not only to {\it describe} the existing data
but also to {\it predict} many bright qualitatively new phenomena.
\end{abstract}

\begin{center}
CONTENTS
\end{center}

\noindent 1. Introduction\\
2. QCD equations\\
3. Comparison with experiment\\
3.1. The energy dependence of mean multiplicity\\
3.2. Difference between quark and gluon jets\\
3.3. Oscillations of cumulant moments\\
3.4. The hump-backed plateau\\
3.5. Difference between heavy- and light-quark jets\\
3.6. Colour coherence in 3-jet events\\
3.7. Intermittency and fractality\\
3.8. The energy behaviour of higher moments of multiplicity distributions\\
3.9. Subjet multiplicities\\
3.10.Jet universality\\
4. Conclusions and outlook\\

\section{Introduction}

Multiparticle production is the main process of very high energy particle
interactions. Studying it, one hopes to get knowledge on the validity of our
general ideas about the structure of the matter at smallest distances, on
new states of matter which could be created at these extreme conditions, 
on the asymptotic properties of strong interactions, on the confinement
of quarks inside hadrons etc.
One should also understand these processes to separate the signals for new
physics from the conventional background, in particular, from the features
explainable in the framework of quantum chromodynamics. This is necessary for
the successful planning of new experiments as well.

With studies of cosmic rays and the subsequent steady increase of energies of
particle accelerators, the interest to multiparticle production processes
increased as well. The theoretical interpretation of these processes passed
through several
stages, in detail described in many monographs and review articles. It evolved
from the statistical and hydrodynamical approaches to the peripheral,
multiperipheral, Regge, parton models and QCD.
Let us stress that all of them are somehow used now (quite often in various
combinations for the description of different stages of the process) in
the computerized Monte Carlo versions available for the detailed comparison with
experimental data.

The ideas of quantum chromodynamics, where quarks and gluons
play a role of partons, nowadays prevail even in the phenomenological models.
Neither these phenomenological models nor the widely used analytical
perturbative QCD approach can evade an important problem
of the transition from partons (quarks, gluons) to the observed particles. This
stage is treated phenomenologically within some hypotheses in any of them. It
introduces additional parameters which, in principle, can give us a hint to
the confinement property, but they sometimes are hard to control and directly
extract.

The whole process of multiparticle production is considered as a jet-like
cascade of consecutive emissions of partons each of which produces the hadrons
observed. Jets from primary quarks were discovered in
$e^{+}e^{-}$-collisions in 1975 with the angular distribution
expected for spin 1/2 quarks. Gluons emitted by quarks at large transverse
momenta can be described by perturbative QCD due to the asymptotic freedom
property according to which the coupling strength in QCD decreases with the
transferred momentum  increasing.
Such processes are used to determine the value of the coupling
strength. However, one can try to proceed to lower transverse momenta when
many jets (and consequently many hadrons) are created. These processes are 
of main concern for this survey.

Main attention will be paid to the process of
$e^{+}e^{-}$-annihilation at high energies, where two point-like particles
collide and the initially created state is 
determined by the time-like quark-antiquark pair whose evolution gives rise
to the production of additional jets. Perturbative QCD has been widely used
for the treatment of these processes (as well as for hard jets
created in the final states of $ep, p\overline{p}$ and similar interactions).
Its predictions and comparison with experimental data are described in the
monographs \cite{1, 2, 3, 4, 5} and recent review papers
\cite{dre1, koch, dgar, kowo, stir}. The application of the QCD ideas to comparatively
soft processes in hadronic and nuclear collisions requires for some modification
of the approach with account for the internal structure of colliding objects
(structure functions) as this is described, e.g., in the surveys 
\cite{glry, lrys, lipa, fstr, wang, schm}.

To be compared with experimental data, the results of quantum chromodynamics
should be presented either by analytical formulas or numerically as computer
calculations according to the Monte Carlo models.

Usually, the Monte Carlo models deal with matrix elements (actually, with
the probabilities) of a process at the parton level plus the hadronization
stage which uses either the string model \cite{agis} or the
fragmentation of hadronic clusters \cite{mawe, mwak}. They
properly account for the energy-momentum conservation laws because all
characteristics of the exclusive process are calculated by computing.
One gets all possible characteristics at a given energy but fails to learn
their asymptotic behaviour. Let us note that many Monte Carlo models have
been proposed which differ by the relative role of the parton and 
hadronization stages and, correspondingly, by the set of adjusted parameters,
which can be interrelated and sometimes are hard to control. Some freedom in
their choice also defines the difference in their predictions.

On the contrary, the analytical approach in QCD pretends to start with
asymptotic values, where the energy-momentum conservation laws can be
neglected in the first approximation, and proceeds to lower energies
accounting for conservation laws,
higher order perturbative and simplified non-perturbative effects.
In the analytical calculations, the perturbative evolution of the jet with the
virtuality of the initial parton degrading due to emission of secondary partons
is terminated at some low scale $Q_0$ of the order of some hundreds MeV for
transverse momenta or virtualities of partons. Some observed variables, e.g.,
such as the thrust or energy flows, are insensitive to this "infrared" cut-off, i.e.
they do not change if collinear or soft partons are additionally emitted.

For others, like the inclusive distributions, the local parton-hadron
duality (LPHD) is assumed \cite{adkt} which
declares that the calculated distributions at the parton level describe the
hadron observables up to some constant factor. This concept originates
\cite{aven, bcma, mtve} from the preconfinement property, i.e. from the
local compensation of colour charges, and, consequently, from the tendency of
quarks and gluons to form the colourless clusters.
Surely, the similarity of the distributions of partons and hadrons substantially
owes to the fact that physics of the colour confinement is governed by rather
soft processes with small momentum transfers.
This works surprisingly well when applied  for comparison of the theoretical
parton distributions with the hadron characteristics observed in experiment. No
phenomenological hadronization model is used in this case. At the same time,
some specific effects, e.g., the Bose-Einstein correlations for the
identical mesons, are out of the scope of this treatment. This is only
admissible if these effects are reasonably small.

Even more amazing look two other features of the perturbative QCD approach: the
probabilistic description and its applicability to the comparatively soft
processes with low transverse momenta\footnote{For more details see
Refs. \cite{5, kowo}}. In the physical
gauge, the leading (at high energy) terms appear as the squared moduli of the 
amplitudes, i.e. as the probabilities.
At high energies, the quantum interference of different amplitudes for the
parton production results in the angular (or, more precisely, transverse
momentum) ordering
of successive emissions of gluons which, in its turn, does not spoil the
probabilistic equations for these processes.

The solutions of these equations
obtained via the modified perturbative expansion sometimes seem to be applicable 
even for rather soft processes where the perturbative expansion parameter is
not sufficiently small 
and, moreover, it is multiplied by some large factors increasing with energy.
Thus, such an approach can be
justified only because some subseries of the purely perturbative expansion 
ordered according to their high energy behaviour are summed first (from here
originates
the term "modified perturbation theory") and then the asymptotic series is cut
off at the proper order.

Surely, the probabilistic treatment is just the 
simplest approach to the whole description of the process valid where the
colour coherence is not important, the non-linearity of the process, induced, e.g.,
by the high density of gluons, and, consequently, the unitarization are not
taken into account etc. If one has to take these effects into account, then
the direct consideration of the traditional Feynman diagrams is preferred
because the influence of these specific effects in terms of the generating
functional technique is hard to treat. Unfortunately, in this case it becomes
more difficult to proceed to the higher order approximations as well.
In this framework of the combined study of the equations for the generating
functionals and of the interference effects implied by Feynman graphs the
perturbative QCD has demonstrated its very high predictive power.

In this paper, we briefly describe the main obtained physical results about the
processes of the production of hadronic jets, mostly in $e^{+}e^{-}$-annihilation
at high energies, without the detailed discussion of any specific particular
technicalities both of the theory and of the experiment. The main goal is to
provide the reader with a short guide  in this vast region without omission,
if this is possible, of main important results and no hiding of newly 
appearing problems. I am deeply sorry if something essential has been omitted
and ask for an excuse from those authors whose papers did not fit the
limited space of this review paper.

For example, what concerns
the theoretical approaches, after their brief description we demonstrate only
the main equations and their solutions with more detailed discussion of the
obtained results, their correspondence to experimental data and physical 
meaning. This would allow the
reader to get quickly acquainted with the general situation and then, following
the numerous cited original papers, more voluminous surveys and monographs, to
study it in more detail if some interest to the particular problem has appeared.
The comparison of theoretical results with experimental data is demonstrated in some
figures which constitute only a small part of those available in the papers
referred to.

Thus this review paper is mainly aimed at those who want
to learn about the achievements and problems in the applications of quantum
chromodynamics to the description of multiparticle production processes at high
energies, especially, those who start their
study of high energy particle physics or specialists in other fields.

\section{QCD equations}

The most general calculational approach to the characteristics of
multiparticle production processes starts from the expression for the generating
functional \cite{kuve, 44}, the equations governing its evolution in the framework of the
considered theoretical scheme and the subsequent solution of these equations.
The generating functional contains the complete information about any
multiparticle process and is defined as
\begin{equation}
G(u, y)=\sum _n\int d^3k_1...d^3k_n u(k_1)...u(k_n)P_n(k_1,...,k_n;
y),     \label{gukk}
\end{equation}
where $P_n(k_1,...,k_n; y)$ is the probability density for the exclusive
production of $n$ particles with momenta $k_1,...,k_n$ at the initial virtuality
(energy), proportional to $\exp (y)$, and $u(k)$ is an auxiliary function. For $u(k)=$const,
one gets the generating function of the multiplicity distribution $P_n(y)$:
\begin{equation}
G(u, y)=\sum _{n=0}^{n=\infty }u^nP_n(y).   \label{guy}
\end{equation}
The variational derivatives of $G(\{u\})$ over $u(k)$
(or usual $u$-derivatives
for constant $u$) provide any inclusive and exclusive distributions (in
particular, the average values and correlators of any rank), i.e. the complete
information about all characteristics of the
process. This information is, however, unavailable unless the equations for 
the evolution of the generating functional are formulated. These equations
depend, naturally, on the theory which has been used for the description of
these processes.

Two equations have been proposed in quantum chromodynamics to describe the
multiparticle production processes in the different kinematical regions. In brief
they are named by the first letters of the names of their authors. They are
the DGLAP equation \cite{glip, apar, dok1} and the BFKL equation
\cite{lip1, klfa, blip}. The DGLAP equation applies to the process of the
evolution of the virtualities (or transverse momenta) of the system.
The BFKL equation deals with this evolution in terms of the longitudinal
momenta. Correspondingly, the DGLAP equations are used for the description
of the development of the partonic jets where the initial highly virtual
time-like parton\footnote{With large positive 4-momentum squared $Q^2>0$.}
evolves, e.g., as in $e^{+}e^{-}$-annihilation at high energies or hard jets
in hadronic processes (see the review papers \cite{dre1, koch, dgar, kowo, stir}).
In its turn, the longitudinal momentum evolution is typical for the
soft\footnote{When the longitudinal momentum transfers $x$ are shared by small
portions; $x\ll 1.$} exchange in the $t$-channel of the interacting particles
which is of the multiperipheral character with some rescatterings
(see the review papers \cite{glry, lrys, lipa, fstr, wang, schm}).
Some proposals for the interpolation between these two regions have been
promoted \cite{ccfm, gust}.

Initially, these equations were derived in the framework of the Feynman diagram
technique by the summation of certain subseries from the whole perturbation series
expansion. Their validity has been proven in the leading and next-to-leading
approximations of the modified perturbation theory (in more detail it will be
described later). Both equations are applicable to the processes
in which the density of the gluon field is comparatively low and the multiparton
interactions and screening may be neglected. They have been successfully applied
to interpretation of many experimental results and for prediction of new
properties of various distributions of secondary particles.

However, these simplified versions of 
both equations have many unsolved problems. They are partly described below in
the case of the DGLAP equation. What concerns the BFKL equation, let us just
mention only one, however important, problem related to the prediction of the
power-like increase of the total cross sections with energy which violates the
unitarity condition and contradicts to the Froissart bound.
According to this bound, the total cross sections of hadronic processes are
allowed to increase with energy $s$ not faster than $\sigma \propto \log^2s$.
In the leading order of the modified perturbation theory, the BFKL equation
predicts the power-like increase of the total cross sections with  very high
exponent $\sigma \propto s^{0.5}$, that strongly contradicts not only to
theoretical restrictions but also to all modern experimental results. Recently,
it was shown \cite{lln, dlip} that the corrections due to the next order terms
are quite large, and even though the power dependence of cross sections on
energy persists, the exponent becomes much lower $\sigma \propto s^{0.17}$.
Such an increase is already much closer to the phenomenological fits of
experimental data at presently available energies which give rise to the
exponent values in the range from 0.08 to 0.12. One can hope that the higher
order corrections will lead to further decrease of the exponent while the
summation of the subseries of the perturbative expansion would be able to avoid
the contradiction with the Froissart condition.

The unified field-theoretical description of both approaches can be
obtained if the proper effective action of the theory 
\cite{mlve, jklw, leon, jmkw, lln, dlip} is constructed. In particular, in
the papers \cite{mlve, jklw, leon, jmkw} apart from the term of the
standard Yang-Mills gluonic action, the effective action contains the term
corresponding to the
non-abelian eikonal interaction of the fastly moving sources with the relevant
component of the gluon field.
Such field-theoretical scheme with the effective action admits the
generalization of these equations to the non-linear case. In the framework of
the Wilson renormalization group approach one can derive the unified functional
QCD equation \cite{jklw, leon} in the leading logarithmic approximation which 
takes into account the higher orders of the parton density fluctuations. It
leads to the system of interconnected equations for the partonic correlators
of an arbitrary order. However, a lot of work is still needed in this direction
as to develop the higher order approximation approach as to understand the role of
the mechanism of the non-linear contributions and the unitarity bounds.

As previously mentioned, this review is mainly devoted to the
$e^{+}e^{-}$-annihilation processes where
the experimental data are most precise and extensive ones. Therefore the
solution of the DGLAP equations for the evolution of partonic jets constitutes
the main content of the theoretical approach adopted here\footnote{In the
forthcoming issues of the Physics-Uspekhi journal, it is intended to publish
the surveys on the effective action, on description of soft processes in
quantum chromodynamics (physics of small $x$ and its relation to the Regge
approach).}.

The general structure of the equation for the generating functional
in QCD describing the jet
evolution for single species partons can be written symbolically as
\begin{equation}
G'\sim \int \alpha _SK[G\otimes G-G]d\Omega.       \label{symb}
\end{equation}
It shows that the evolution of the functional $G$ indicated by its derivative
$G'$ over the evolution parameter (the transverse momentum or the virtuality)
 is determined by the cascade process of the production of two partons
by a highly virtual time-like parton (the term $G\otimes G$) which provides
new partons in the phase space volume considered $d\Omega $ and by the
escape of a single parton ($G$) from a given phase space region.

Therefore
this equation contains terms corresponding to inflow and outflow of partons.
In fact, it can be interpreted as the kinetic equation with the collision
integral in the right hand side. The weight factors are determined by the
coupling strength $\alpha _S$ and the splitting function $K$ which is defined
by the interaction Lagrangian. The integral runs over all internal
variables, and the symbol $\otimes $ shows that the two created partons share the
momentum of their parent. The initial condition for equation (\ref{symb})
is defined by the requirement for the jet to be created by a single initial
parton, i.e., by 
\begin{equation}
P_n=\delta _{n1}; \;\;\;\;\;\;\;\;\; G_0=u(k).
\end{equation}
It is clear from this formula that we have to deal with the non-linear
integro-differential probabilistic equation with shifted arguments in
the $G\otimes G$ term under the integral sign.

For quark and gluon jets, one writes down the system of two coupled equations.
Their solutions give all characteristics of quark and gluon jets and allow for
the comparison with experiment to be done. Let us write now them down explicitly 
for the generating functions:
\begin{eqnarray}
&G_{G}^{\prime }&= \int_{0}^{1}dxK_{G}^{G}(x)\gamma _{0}^{2}[G_{G}(y+\ln x)G_{G}
(y+\ln (1-x)) - G_{G}(y)] \nonumber \\ 
&+&n_{f}\int _{0}^{1}dxK_{G}^{F}(x)\gamma _{0}^{2}
[G_{F}(y+\ln x)G_{F}(y+\ln (1-x)) - G_{G}(y)] ,   \label{50}
\end{eqnarray}
\begin{equation}
G_{F}^{\prime } = \int _{0}^{1}dxK_{F}^{G}(x)\gamma _{0}^{2}[G_{G}(y+\ln x)
G_{F}(y+\ln (1-x)) - G_{F}(y)] ,                                   \label{51}
\end{equation}
where $G^{\prime }(y)=dG/dy ,$
$y=\ln (p\Theta /Q_0 )=\ln (2Q/Q_{0})$ is the evolution parameter, defining
the energy scale,
$p$ is the initial momentum, $\Theta $ 
is the angle of the divergence of the jet (jet opening angle), assumed here to be 
fixed, $Q$ is the jet virtuality,  $Q_{0}=$ const , 
$ n_f$ is the number of active flavours,
\begin{equation}
\gamma _{0}^{2} =\frac {2N_{c}\alpha _S}{\pi } ,                \label{52}
\end{equation}
the running coupling constant in the two-loop approximation is
\begin{equation}
\alpha _{S}(y)=\frac {2\pi }{\beta _{0}y}\left( 1-\frac {\beta _1}
{\beta _{0}^{2}}\cdot \frac {\ln 2y}{y}\right)+O(y^{-3}), \label{al}
\end{equation}
where
\begin{equation}
 \beta _{0}=\frac {11N_{c}-2n_f}{3}, \;\;\;\;\;\;
 \beta _1 =\frac {17N_c^2-n_f(5N_c+3C_F)}{3},
 \label{be}
\end{equation}
the labels $G$ and $F$ correspond to gluons and quarks,
and the kernels of the equations are
\begin{equation}
K_{G}^{G}(x) = \frac {1}{x} - (1-x)[2-x(1-x)] ,    \label{53}
\end{equation}
\begin{equation}
K_{G}^{F}(x) = \frac {1}{4N_c}[x^{2}+(1-x)^{2}] ,  \label{54}
\end{equation}
\begin{equation}
K_{F}^{G}(x) = \frac {C_F}{N_c}\left[ \frac {1}{x}-1+\frac {x}{2}\right] ,   
\label{55}
\end{equation}
$N_c$=3 is the number of colours, and $C_{F}=(N_{c}^{2}-1)/2N_{c}
=4/3$ in QCD. The asymmetric form (\ref{53}) of the three-gluon vertex can be 
used due to the symmetry properties of the whole expression. If one puts
$n_f=0$ in equation (\ref{50}) and omits equation (\ref{51}), then one
gets the equation of gluodynamics briefly described above.
The auxiliary variable $u$ has been omitted in the generating functions for
simplicity.

The typical feature of any field theory with a dimensionless coupling constant
(quantum chromodynamics, in particular) is the presence of the singular
terms at $x\rightarrow 0$ in the kernels (\ref{53}), (\ref{55}) of the
equations. They imply the
uneven sharing of energy between newly created jets and play an important role
in the jet evolution giving rise to its more intensive development compared
with the equal proportions (nonsingular) case.

Let us note that these equations can be transformed into the linear
equations for the moments
of multiplicity distributions (see subsections 3.1 and 3.3).
For the running coupling strength (\ref{al}), they have been solved only in the
perturbative theory approach. The systematical method, proposed in \cite{13},
consists in using the Taylor series expansion
of the expressions under the integral sign. It results in the modified
perturbative expansion of the physically measurable quantities.
Namely such solutions are usually considered when the analytical formulas
are compared with experimental results.

At the same time, these equations can be exactly solved \cite{21, dhwa}
if the coupling strength is assumed fixed, independent of $y$, i.e.
$\alpha _S$=const instead of (\ref{al}). This becomes possible because of the
scaling property leading to the power-like behaviour of the average multiplicity
and all higher moments for the fixed coupling case. This behaviour differs from
the dependences obtained in the case of the running coupling strength (see
subsection 3.1). Therefore, the slow logarithmical decrease of this strength
with transferred momenta increasing is crucial for the development of the
parton cascades.

Even though the system of equations (\ref{50}), (\ref{51}) is physically
appealing, it is not absolutely exact, i.e., it is not derived from the first
principles of quantum chromodynamics. One immediately notices this since,
for example, in these equations there is no
four-gluon interaction term which is contained in the lagrangian of QCD.
This interaction should, probably, correspond to the contribution to the
integral term of the equations proportional to the product of three generating
functions with a corresponding weight factor.
Such a term does not contribute the singularity to the kernels and its
omission is justified in the lowest order approximations.
There were no attempts to take this term into account in the higher order
perturbative expressions.

Nevertheless, the modified series of
the perturbation theory with three-parton vertices is well
reproduced by such equations up to the terms including two-loop and
three-loop corrections. As shown in Ref. \cite{5}, the neglected terms would
contribute at the level of the product of, at least, five generating functions.
The physical interpretation of the corresponding Feynman graphs would lead to the
treatment of the 'colour polarizability' of jets.

These equations are justified up to some approximation of the modified 
perturbation theory (see below), because they only include those Feynman
diagrams, where the gluons have strongly ordered transverse momenta, and are
not close to the kinematical limit. Apart from this,
 they take into account the non-perturbative
effects (e.g., the properties of the QCD vacuum) only in a simplified manner
by the direct cut-off of the cascade evolution at some virtuality $Q_0$. 
In principle, the effective infrared-safe coupling constant (without the Landau
pole) \cite{shir} may be used as the
substitute for the phenomenological parameter $Q_0$. It must be universal 
for different processes and tend to a constant limit at low virtualities. 
The constant average value of the coupling strength in this region has been
used in papers \cite{marc, dmwe} dealing with the non-perturbative corrections.
However, the behaviour of the coupling strength is not the only non-perturbative
effect. Therefore, we will use the more traditional perturbative approach. 
The non-perturbative effects in the three-jet events have been studied in more
detail in the papers \cite{bdmz}.

There are some problems also with
the definition of the evolution parameter, with preasymptotic corrections etc.
 For example, the lower and upper limits of integration over $x$ in Eqns
 (\ref{50}), (\ref{51}) are constant and correspond to their asymptotic
values. In reality, they vary in the preasymptotic region. Their form is
determined by the restriction imposed on the transverse momentum which is given
by the inequality
\begin{equation}
k_t = x(1-x)p\Theta ^{\prime } > Q_0/2.   \label{ktli}
\end{equation}
This condition originates from the requirement that the formation time of a
gluon  ($t_{form}\sim k/k_{t}^{2}$) must be less than its hadronisation time
($t_{had}\sim kR^{2}\sim k/Q_{0}^{2}$). It should be imposed  for the
perturbative QCD to be 
applicable. This leads to the requirement that the arguments of the generating
functions in Eqns (\ref{50}), (\ref{51}) should be positive. Therefore,
we must integrate in Eqns (\ref{50}), (\ref{51}) over 
$x$ from $\exp (-y)$ to $1-\exp (-y)$. However these limits tend to 0 and 1 at 
high energies ($y\rightarrow \infty $). The omitted contributions decrease in
the power-like manner with energy increasing.
That is why it seems reasonable to learn more about the 
solutions of equations (\ref{50}), (\ref{51}) near the asymptotic region
taking first into account the perturbative (logarithmically decreasing) 
corrections, and only then take the neglected power-like terms into account
as further corrections to these solutions.

Moreover, this cut-off of the limits of integration is of physical importance.
With the limits equal to $\exp (-y)$ and $1-\exp (-y)$, the partonic cascade
terminates at the perturbative level $Q_0/2$ as is seen from the arguments of
the generating functions in the integrals. With the limits equal to 0 and 1,
one extends 
the cascade into the non-perturbative region with low virtualities
$Q_1\approx xp\Theta /2$ and $Q_2\approx (1-x)p\Theta /2$ less than $Q_0/2$.
Namely this region contributes terms of the order of $\exp (-y)$, power-suppressed
in energy. It is not clear whether the equations and LPHD hypothesis are valid
down to some $Q_0$ only or the non-perturbative region can be included as well.

Some approximations are used to solve these equations with the running coupling
strength. The Taylor series expansion leads \cite{13} immediately, as will be
shown below, to the perturbation theory series in the exponent of the physically
measurable quantities. It implies the summation of some specific subseries of
the purely perturbative expansion in the coupling strength for this particular
characteristics. This justifies the term of the "modified perturbative expansion"
attached to this procedure. Moreover, the expansion parameter is not $\alpha _S$
itself but its square root $\gamma _0$.
The asymptotic results (for extremely high energies)
are obtained in the so-called double-logarithmic (DLA or DLLA if it is called
as double-leading-logarithmic) or leading order (LO)
approximation where the terms $(\alpha _S\ln ^2s)^n$ are summed. Here $s$ is 
the cms energy squared. The emitted gluons are assumed to be so soft that the
energy-momentum conservation is neglected.

The corrections accounting for
conservation laws in the $G\otimes G$ term, i.e., the shift in their arguments 
in equations (\ref{50}), (\ref{51}),
as well as the higher order terms in the weight $\alpha _SK$ (in particular,
the non-singular terms of the kernels $K$ and the dependence of $\alpha _S$ on
the transverse momentum $k_t$ (\ref{ktli})) appear in the
next-to-leading (NLO or MLLA - modified leading logarithmic,
NLLA - next-to-leading logarithmic approximation) and 
higher (2NLO, 3NLO, ...) orders. Formally, these equations have been proven only for
the next-to-leading (NLO) order of the modified perturbative QCD. However, one
can try to consider them as the kinetic equations and solve them in higher orders
with a hope to get from the obtained solutions and their comparison with
experiment any indications on the role of the adopted assumptions.

To have a
guide to the uncertainties at higher orders of perturbation theory, one can
try to generalize these equations by including the abovementioned effects in a
more rigorous way than it is usually implied. However, there is no unique way
of doing it. 
Also, it is not clear how one may modify them to include the non-perturbative
effects, the colour coherence, the non-linearity at high densities etc, even 
though the very preliminary phenomenological versions have been proposed
\cite{123, 124, ccfm, 125, 127, 128, eden, gust, dede}. The simplest proposal
would be to
compare two alternative evolution equations which use somewhat different
assumptions leading to different higher order contributions as has been
demonstrated in \cite{eden, dede}.

At the end of the section, let us mention that no direct solution of
equations (\ref{50}), (\ref{51}) for the generating
functions depending on both variables $u$ and $y$ has been obtained. Such
solutions would be of interest for finding the location and the strength of 
the singularities of $G$ in the complex plane $u$. This is important in connection
with the behaviour of the moments of the distribution (see formulas (\ref{4}),
(\ref{5}) below) and with some analogies from the statistical physics, where
these singularities would indicate the phase transition point \cite{lyan, ylee}.
The corresponding singularities have been found only in the lowest order of 
the modified perturbation theory.
It occured that they are located at the point $u=1+z_0$, where
$z_0=C/\langle n\rangle  (C\approx 2.552) $ tends to 0 at high energies. The
singular part of $G$ at this location looks like
\begin{equation}
G(u, y)=\frac {2z_0^2}{(u-1-z_0)^2}+\frac {2z_0}{u-1-z_0}-\frac {2}{3}\log
\frac {z_0+1-u}{z_0}+O(1).   \label{sing}
\end{equation}
The structure of singularities is rather complicated. They are close
to the point $u=1$ where all moments are calculated.
Moreover, with the energy increase they move closer to this point
and coincide with it in asymptotics. There is, however, no special reason to
worry about it since the generating function should be equal to 1 at this point
according to its definition, and, therefore, all singularities must cancel
somehow there. This is seen from the above expression as well.
Nonetheless, this indicates the completely different structure of
the singularities which one obtains in the lowest order perturbative expressions 
compared with the final result.

In experiment, one always has to cut off the sum over the multiplicity in the
definition of the generating function (\ref{guy}) at some maximum multiplicity
$n_{max}$, defined either by the energy-momentum conservation laws or by the
definite conditions of a particular experiment. Then the generating function
becomes a polynomial of the order of $n_{max}$ with positive coefficients and,
therefore, possesses just this number of the complex conjugated roots.
Thus, one can find out the location of the singularity from experimental data
only by increasing energy (and, consequently, the value of $n_{max}$) and
following the evolution of their locations. According to the papers
\cite{lyan, ylee}, the complex-conjugated roots of the partition function of
the grand canonical ensemble pinch the real axis of $u$ at the limiting
transition to the infinite volume just at the singularity location. The
behaviour of the same type has been demonstrated in multiparticle production
experiments for increasing $n_{max}$ (in more detail, see Ref. \cite{dgar}).

\section{Comparison with experiment}

Let us turn now directly to the comparison of the theoretical results obtained
with available experimental data. The main bulk of the data is provided by
$e^{+}e^{-}$-processes at the $Z^0$ energy. Many results can be found in the
compilation \cite{80}.

In experiment, one registers the particles created at
the final stage, mostly, hadrons. Then the crude data are corrected for
the effectiveness of the detectors, possible radiation of photons before 
the collision etc.  At the same time, as has been mentioned before, the theory
provides us with knowledge of the properties of quark and gluon jets
and of the distributions of partons created during their evolution. Therefore,
the problem of the correspondence of the theoretical results to the
experimental data arises. One must separate the jets in an adequate manner
and compile the "dictionary" for the translation from the parton to
hadron "language".

In $e^{+}e^{-}$-annihilation processes, the quark-antiquark pair is first
created. Therefore, the measurement of the multiplicity of the quark jet
is of no problem because it is just twice smaller than the total multiplicity
and equals to the multiplicity in one of the hemispheres. In terms of the
generating functions this can be expressed by the relation
\begin{equation}
G_{e^+e^-}\approx G_F^2.
\end{equation}

To get the analogous
results about the gluon jets, one should, however, have the access to the
pairs of the gluon jets created by a colour-neutral source. This is necessary for
the complete correspondence to the theoretical definitions. Unfortunately,
such a separation of these events is possible with satisfactorily high precision
only either in the decay processes
$\Upsilon \rightarrow \gamma gg\rightarrow \gamma +$hadrons \cite{7} or in the
rare collisions leading to the almost parallel heavy quark and antiquark in the
same hemisphere with the gluon in the opposite hemisphere \cite{139, 140, 9}.
The high experimental statistics at Z$^0$ energy allows to do this. At other
energies, where the number of registered events is much smaller, the separation
of gluon jets with the help of methods, which do not depend on the chosen
algorithm (the so-called "unbiased jets"), became possible quite recently
after application of the special analysis of the two- and three-jet events
according to two variables - the transverse momentum and energy - proposed
in \cite{139, egus, egkh} and used in \cite{488} (see the subsection 3.6).
The separation of the quark and gluon jets is often done with the help of
some special algorithms\footnote{The most popular one is the method of the jet
separation according to their relative transverse momentum called $k_t$- or
Durham-algorithm \cite{231}}. Their use is sometimes not completely identical
to the theoretical requirements (the so-called "biased jets"). Then one has
to rely only on the comparison with 
the results of those Monte Carlo models where the applied algorithm has been 
taken into account. The choice of the algorithm is by itself determined by its
physical reliability, by the convenience for the analytical estimates and by the
role of the hadronization corrections in these estimates.

What concerns the hadronization stage, i.e., the transition from partons to
hadrons, the Monte Carlo models use various phenomenological approaches
mentioned briefly above. In the analytical calculations, the variation of the
parameter $Q_0$ in the combination with the hypothesis of the local
parton-hadron duality usually plays the role of the "dictionary" and leads to
reasonable results.

\subsection{The energy dependence of mean multiplicity}

The equations for the average multiplicities in jets are obtained from
the system of equations (\ref{50}), (\ref{51}) by expanding the generating
functions in the power series of $u-1$ and keeping only the terms
with $q$=0 and 1 with account of the definition of the average multiplicity
$\langle n\rangle $ as
\begin{equation}
\frac{dG}{du}\vline _{u=1}=\sum nP_n=\langle n\rangle.
\end{equation}
Finally, one gets the linear integro-differential equations with the shifted
arguments under the integral sign for the average multiplicities.
They read
\begin{eqnarray}
\langle n_G(y)\rangle ^{'} =\int dx\gamma _{0}^{2}[K_{G}^{G}(x)
(\langle n_G(y+\ln x)\rangle +\langle n_G(y+\ln (1-x)\rangle -\langle n_G(y)
\rangle ) \nonumber  \\
+n_{f}K_{G}^{F}(x)(\langle n_F(y+\ln x)\rangle +\langle n_F(y+
\ln (1-x)\rangle -\langle n_G(y)\rangle )],  \label{ng}
\end{eqnarray}
\begin{equation}
\langle n_F(y)\rangle ^{'} =\int dx\gamma _{0}^{2}K_{F}^{G}(x)
(\langle n_G(y+\ln x)\rangle +\langle n_F(y+\ln (1-x)\rangle -\langle n_F(y)
\rangle ).   \label{nq}
\end{equation}

From here, by solving these equations, one can learn about the energy evolution
of the ratio of average multiplicities between gluon and quark jets $r$ and
of the QCD anomalous dimension $\gamma $
(the slope of the logarithm of average multiplicity in the gluon jet) defined as
\begin{equation}
r=\frac {\langle n_G\rangle }{\langle n_F\rangle }\; ,\;\;\;\;\; \;\;\;
\gamma =\frac {\langle n_G\rangle ^{'}}{\langle n_G\rangle }
=(\ln \langle n_G\rangle )^{'}\; .  \label{def}
\end{equation}
They have been represented
by the perturbative expansion at large energy (or large $y$) as
\begin{equation}
\gamma = \gamma _{0}(1-a_{1}\gamma _{0}-a_{2}\gamma _{0}^{2}-a_3\gamma _0^3)+O(\gamma _{0}^{5})
 , \label{X}
\end{equation}
\begin{equation}
r = r_0 (1-r_{1}\gamma _{0}-r_{2}\gamma _{0}^{2}-r_3\gamma _0^3)+O(\gamma _{0}^{4})
.  \label{Y}
\end{equation}
Using the Taylor series expansion \cite{13} of $\langle n\rangle $ at large $y$
in Eqns (\ref{ng}), (\ref{nq}) with (\ref{X}), (\ref{Y}) and equating the terms 
of the same order in $\gamma _0$ in both sides, one gets the
coefficients $a_i,\, r_i$ shown in the Table 1 ($r_0=9/4$).

Table~1

\begin{tabular}{|c|c|c|c|c|c|c|}
\hline
$n_f$ & $r_1$ & $r_2$ & $r_3$ & $a_1$ & $a_2$ & $a_3$\\
\hline
3 &  0.185 & 0.426 & 0.189   & 0.280 & - 0.379 & 0.209\\
\hline
4  &  0.191 & 0.468 & 0.080  & 0.297 & - 0.339 & 0.162 \\
\hline
5 &    0.198 & 0.510 &  -0.041  & 0.314 & - 0.301 & 0.112\\
\hline
\end{tabular}

The parameter $\gamma $ determines the exponent of 
the mean multiplicity of the gluon jet
\begin{equation}
\langle n_G \rangle = \exp (\int ^{y}\gamma (y\prime )dy\prime ),   \label{57}
\end{equation}
and therefore its perturbative expansion in terms of $\gamma _0$ corresponds in
each particular order in $\gamma _0$ to the summation of some subseries in the
purely
perturbative expression for the average multiplicity, i.e., to the modified
perturbation theory for this particular physical quantity. The choice of the
gluon jet is related, first of all, with the tradition to study in the beginning
the limiting case of gluodynamics where $n_f=0$, i.e., quarks and,
correspondingly, equation (\ref{nq}) are not cosidered. This is justified
also because in the lowest order approximations (the NLO corrections including)
the energy dependence of average multiplicities in quark and gluon jets
do not differ (for more detail, see next subsection).

One of the most spectacular predictions of QCD states that in the leading order
approximation (i.e., asymptotically), where $\gamma =\gamma _0$, average
multiplicities should increase with energy \cite{mu1, dfkh, bcmm} like
$\exp [2c\sqrt {\log s}]$ with the theoretically calculable value of $c$.
This behaviour is just in between the power-like and logarithmical
dependences correspondingly predicted by the hydrodynamical and multiperipheral
models\footnote{
It is not excluded, however, that the non-linear corrections which appear, say,
due to the high gluonic density can unitarize this approximation and give rise
to the logarithmic increase of the mean multiplicity in the asymptopia
\cite{glry, jklw, jmkw}.}.
Next-to-leading order results account for the term with $a_1$ in Eqn. (\ref{X})
\cite{web1, dktr, cdfw} and contribute the logarithmically decreasing
factor to this behaviour. Namely these two terms determine the main energy
dependence of average multiplicities, and they are the same for quark
and gluon jets. The higher order terms do not practically
change this dependence \cite{dg, cdnt8}:
\begin{eqnarray}
\langle n_{G}\rangle=Ky^{-a_{1}c^2 }\exp ( 2c\sqrt y+
\frac {c}{\sqrt y}[2a_2c^2+\frac {\beta _1}{\beta _{0}^{2}}(\ln 2y+2)]
\nonumber \\
 +\frac {c^2}{y}[a_3c^2-\frac {a_1\beta _1}{\beta _{0}^{2}}(\ln 2y +1)]),
 \label{mean}
\end{eqnarray}
where $c=(4N_c/\beta _0)^{1/2}$.
The gluodynamics expressions can be obtained from this formula for $n_f=0$
taking into account it also in the analytical expressions for $a_i$ given
in \cite{cdnt8}.

The fitted parameters in the final expression are the overall constant
normalization factor $K$ which is defined by the confinement\footnote{That is
why the lower limit of the integration over $y'$ in formula (\ref{57}) is
not fixed.} and the scale parameter $Q_0=2\Lambda $. The $e^{+}e^{-}$-data
in the energy interval from the $\Upsilon $-resonance to LEP-2 (i.e.,
approximately from 10 to 200 GeV) are well fitted by such an
expression as seen in Fig. 1. The dotted line corresponds to the fit with two
adjustable parameters. The dashed line shows the fit by the Monte Carlo model
HERWIG. Other lines determined by the difference between gluon and quark jets
will be explained in next subsection.

Let us note here that the expansion parameter
$\gamma $ is rather large at present energies ranging from 0.4 to 0.5.
The obtained expressions are valid at high energies and do not pretend to
describe the energy behaviour of mean multiplicities near threshold.
The computer solutions \cite{lo2, lo1} of equations (\ref{50}), (\ref{51}) 
give rise to the satisfactorily precise agreement with experimental data even
at rather low energies.

Equations (\ref{ng}), (\ref{nq}) can be solved exactly in the case of the
fixed coupling constant \cite{21, dhwa} because of the scaling property
according to which the relation
\begin{equation}
\frac{\langle n(y+\ln x)\rangle }{\langle n(y)\rangle }=x^{\gamma }
\end{equation}
is valid for $\gamma _0$=const and integro-differential equations
(\ref{ng}), (\ref{nq}) are reduced to the system of easily solvable
linear algebraic equations. The energy increase of mean multiplicity
becomes power-like because the exponent acquires the general structure of
the type of $\gamma _0\ln s$.

\subsection{Difference between quark and gluon jets}

The system of two equations for quark and gluon jets predicts that
asymptotically the energy dependence of mean multiplicities for them should
be identical. Moreover, this coincidence is exact in the next-to-leading
approximation of the modified perturbation theory as well. Higher order
correction, even though violating this beautiful feature, are comparatively
weak in the functional dependence on energy. Namely this explains the initial
success in the description of the energy dependence of average multiplicities
in $e^{+}e^{-}$-annihilation in the framework of gluodynamics.

The absolute normalization is not fixed. However, the relative
normalization, as given by $r$ (\ref{Y}), is calculable.
The gluon jets are more "active" than the quark jets so that the ratio
$r=\langle n_G\rangle /\langle n_F\rangle $ of average multiplicities between
gluon and quark jets should tend at high energies \cite{brgu}
to the ratio of the Casimir operators $C_A/C_F=9/4$.

Once again, by comparison with experiment this prediction shows how far are we
now from the true asymptotics. Even though this prediction obtained in the
leading order (LO) approximation
is valid qualitatively, its quantitative value is still rather far from
experimental ones where this ratio is about 1.5 at $Z^0$ energy and even smaller
at lower energies. The higher order\footnote{The difference in the definition
of the "order" for the anomalous dimension $\gamma $ and for the ratio of mean
multiplicities $r$ is described in Ref. \cite{cdnt8}. In particular, the term
$r_3$ in $r$ should be actually considered as 4NLO correction since it is
added to the still uncalculated term $a_4$ in the higher order expression
for the quark jet anomalous dimension $\gamma _F$ and, consequently, influences
its multiplicity.} terms \cite{43, gmue, mweb, web1, cdnt8, 41} are rather
important just for this parameter. They have been calculated now up to
3NLO (or 4NLO; see the footnote) terms (see Table~1) and improve the agreement
approaching the experimental value with an accuracy about 15$\%$ (see Fig. 2
where the old notation for the approximation order has been used). Let us stress
again that just the ratio of mean multiplicities $r$
(but not the energy behaviour of any of them, 
$\langle n_G\rangle $  or $\langle n_F\rangle $, separately) is most sensitive
to these corrections. This is because the main LO and NLO terms cancel there
and decline from the constant value $r_0$ is governed by the higher order
terms. This decline can provide a guide to further generalizations of the
equations and to the proper account of the nonperturbative contributions.

Surely, the higher order terms change slightly also the energy behaviour
of multiplicities for quark jets compared to gluon jets\footnote{The value
$\gamma $ in (\ref{57})
is replaced there by $\gamma _F=\gamma -r'/r$, which differs from $\gamma $ only
in higher orders as is easily estimated if one recalls that
$r'\sim \gamma _0'\sim \gamma _0^3$ (see below formula (\ref{r2r1})).}
as observed in experiment. However, the simultaneous fit of quark and gluon
jets with the same set of fitted parameters even in the framework of 3NLO
approximation is still
not very accurate as is seen from the shaded area in Fig. 1.
This area demonstrates the attempt of such a description with the data on 
gluon jets divided by the theoretically (analytically) calculated value
of the ratio $r$.
Its failure shown by the shift of the shaded area compared to the experimental
points and by its large width is again due to the insufficiently precise description
of the ratio $r$. The agreement with experimental multiplicity in $e^{+}e^{-}$
is restored either in the case of the normalization according to gluon jets with
the only adjusted parameter $Q_0=2\Lambda $ (the solid line in Fig. 1) or
in the case with both parameters fitted (the dotted line).

The computer solution of the equations \cite{lo1} gives rise to very good
agreement with experiment at the $Z^0$ resonance and leaves a rather small
difference about 20$\%$ even at so low energies as the mass of the $\Upsilon $.
In the analytical approach, the quite good agreement on the ratio $r$ at the
energy of the $Z^0$ has been achieved when the equations are modified to
account for the phase space limitations imposed by energy-momentum conservation
in the dipole cascade picture of
the string approach \cite{eden}. However, some problems arise for higher moments
of the multiplicity distribution in such an approach \cite{dede}.

Equation (\ref{Y}) states that the ratio $r$ increases with energy
increasing due to the decrease of $\gamma _0$ (the asymptotic freedom) but 
its slope becomes smaller. Experimental values also increase with energy
as seen from Fig. 2. There are common tendencies and good qualitative agreement
between the results of analytical calculations and experiment on the value of
the ratio $r$. However,
the quantitative statements are valid only within the 15 - 25$\%$ accuracy.

More detailed characteristics of jets behaviour can be obtained from the slope
and the curvature of energy dependences of multiplicities \cite{drsl, cdnt8}.
Some experimental data about it have been recently obtained \cite{488}.

The slope of the energy dependence of the ratio of average multiplicities
between gluon and quark jets $r$, i.e., the derivative $r'$ ("the slope of
the ratio"), decreases
with the energy increase according to the expansion (\ref{Y}) as $\ln ^{-3/2}s$.
It is very sensitive to the higher order corrections \cite{drsl, cdnt8}, first
of all, due to the large value of the ratio $r_2/r_1$:
\begin{equation}
r'=Br_0r_1\gamma _0^3[1+\frac {2r_2\gamma _0}{r_1}+(\frac {3r_3}{r_1}+
B_1)\gamma _0^2+O(\gamma _0^3)],  \label{r2r1}
\end{equation}
where $B=\beta _0/8N_c$, $B_1=\beta _1/4N_c\beta _0$. Therefore its use for
comparison with experiment does not seem to be informative enough.

At the same
time, such corrections are partly cancelled in "the ratio of the slopes"
$r^{(1)}=\langle n_G\rangle '/\langle n_F\rangle '$
for gluon and quark jets \cite{drsl, cdnt8}.
Thus, the ratio of the slopes must
depend on energy much weaker. This has been observed in experiment. The first
comparison \cite{dgar} of the theoretical estimates of this ratio
\cite{desl, opsl} with experiment has shown quite good agreement, however,
still within rather large undefiniteness in experimental results. 
Recent more accurate separation of unbiased gluon jets \cite{488}
made it possible to get the values not only the ratios $r$ and $r^{(1)}$ at
energies from $\Upsilon $ to Z$^0$ (including the intermediate ones) but also
the values of the ratio of the second derivatives
$r^{(2)}=\langle n_G\rangle ''/\langle n_F\rangle ''$. The asymptotic values
for all of them are predicted equal 2.25. The corresponding values at the
energy of the hadronic component in $gg$-system 30 GeV are
$r=1.422\pm 0.051, r^{(1)}=1.761\pm 0.071, r^{(2)}=1.92\pm 0.13$ and at 80 GeV
$r=1.548\pm 0.042, r^{(1)}=1.834\pm 0.09, r^{(2)}=2.04\pm 0.14$. The error
bars are mainly determined by the systematic errors. These results are in 
agreement with the analytical QCD predictions \cite{drsl, cdnt8} that at
present energies $r^{(2)}$ should be closer to its asymptotic value of 2.25
than $r^{(1)}$, and $r^{(1)}$ closer to 2.25 than $r$, i.e. the inequalities
$r<r^{(1)}<r^{(2)}$ are valid. The experimentally obtained values of the
ratios of slopes and curvatures of energy dependence of gluon and quark jets
and their comparison with some theoretical calculations are demonstrated
in Figs. 3 and 4.

One can hope that the asymptotic values of the multiplicities ratio $r$
can be approached closer if the soft particles are only considered with energies
much lower than the total energy of the jet \cite{klo1, klo2}. This effect
has been really noticed but it is not very strong. The ratio $r$ for particles 
with low momenta is about 1.8 \cite{soft}.

It is interesting that in the Monte
Carlo model HERWIG accurately reproducing this ratio for soft hadrons the
corresponding ratio for soft partons occurs exactly equal to 2.25. Thus the role
of the hadronization stage is very important for the ratio $r$ in this
particular model. Let us note that this situation differs from the case
where all momenta are averaged as demonstrated in Fig. 2 with the lines of the 
HERWIG model practically indistinguishable for partons and particles. This is
related to the strong difference of the spectra which are rather soft for
hadrons and much harder for partons in this model (for more detail, see Ref.
\cite{dgar}). The direct use of the
local parton-hadron duality is, therefore, impossible here because of the
large value of the adopted cut-off parameter $Q_0$ indicating the early finish
of the parton stage\footnote{I am grateful to Bill Gary for the discussion of
this problem.}

The parton stage in the evolution of the jet of final hadrons can be separated
in a following way. In the HERWIG model it is assumed to finish at
$Q_0$=0.8 GeV. Thus it corresponds to the rather large transverse momenta.
Therefore one should also choose the hadrons with large transverse momenta.
Analysis of the data about the comparatively soft hadrons ($p<4$ Ē'/$c$) with
rather large transverse momenta $p_t>$0.8 Ē'/$c$ have led to the result
\cite{soft} for the ratio $r=2.29\pm 0.017$ in this region which agrees with
asymptotic predictions. In its turn, this result could agree with the
computer solutions of equations (\ref{50}), (\ref{51}) and with the
local parton-hadron duality hypothesis only if one adopts that the evolution of
the "partonic" cascade in the computer calculations to the region $Q_0 < 0.8$ GeV
plays the same role as the "hadronization" stage in the HERWIG model, where
such an evolution is discarded.

The widths of the multiplicity distributions differ for quark and gluon jets,
the former being somewhat wider. Qualitatively, QCD describes this tendency
but quantitative estimates are rather uncertain yet as is discussed in more
detail in subsection 3.8.

\subsection{Oscillations of cumulant moments}

The shape of the multiplicity distribution can be described by its higher
moments related to the width, the skewness, the kurtosis etc. Equations
for the moments of the distributions of the $q$-th rank are obtained from
the system of equations (\ref{50}), (\ref{51}) by comparing the terms
of the type $(u-1)^q$ in both sides of equations. The $q$-th
derivative of the generating function corresponds to the factorial moment
$F_q$, and the derivative of its logarithm defines the so-called cumulant
moment $K_q$. The latter ones describe the genuine correlations
in the system irreducible to the products of lower order correlations
(this recalls the connected Feynman graphs).
\begin{equation}
F_{q} = \frac {\sum_{n} P_{n}n(n-1)...(n-q+1)}{(\sum_{n} P_{n}n)^{q}} =
\frac {1}{\langle n \rangle ^{q}}\cdot \frac {d^{q}G(z)}{du^{q}}\vline _{u=1}, 
\label{4}
\end{equation}
\begin{equation}
K_{q} = \frac {1}{\langle n \rangle ^{q}}\cdot \frac {d^{q}\ln G(z)}{du^{q}}
\vline _{u=1}. \label{5}
\end{equation}
These moments are not independent. They are connected by definite relations
which can easily be derived
from their definitions in terms of the generating function:
\begin{equation}
F_{q} = \sum _{m=0}^{q-1} C_{q-1}^{m} K_{q-m} F_{m} .              \label{11}
\end{equation}
They are nothing other than the relations between the derivatives of a function
and of its logarithm at the point where the function itself equals 1. Here
\begin{equation}
C_{q-1}^{m} = \frac {(q-1)!}{m!(q-m-1)!} = \frac {\Gamma (q)}{\Gamma (m+1)
\Gamma (q-m)} = \frac {1}{mB(q,m)}     \label{12}
\end{equation}
  are the binomial coefficients, and $\Gamma $ and $B$ denote the gamma-  and
 beta-functions, correspondingly. Thus there are only numerical
  coefficients in recurrence relations (\ref{11}) and the iterative
solution (well-suited for computer calculations) reproduces all cumulants if
the factorial moments are given, and vice versa. In that sense, cumulants
and factorial moments are equally suitable for analysis of multiplicity
distributions. The relations for the low ranks are
\begin{eqnarray}
F_{1}&=&K_{1}=1, \nonumber  \\
F_{2}&=&K_{2}+1, \nonumber  \\
F_{3}&=&K_{3}+3K_{2}+1.   \label{12a}
\end{eqnarray}

The Taylor series expansion of equations (\ref{50}), (\ref{51}) gives rise to
the term $G(y)$ in the right-hand side. Dividing by it the left-hand side $G'(y)$
one gets the derivative over $y$ of the logarithm of the generating function.
The gluodynamics equation in the leading order approximation can be reduced
to the differential equation of the second order:
\begin{equation}
(\ln G(y))''=\gamma _0^2(G(y)-1); \;\;\;\; G(0)=u, \;\;\; G'(0)=0. \label{dla}
\end{equation}
After that, the derivatives over $u$ of both sides considered at $u=1$ lead in
a natural way to the
prediction \cite{13, 21, 41} concerning the behaviour of the ratio $H_q=K_q/F_q$.
At asymptotically high energies, this ratio as a function of the rank $q$
is predicted to behave\footnote{This is clearly seen from equation (\ref{dla})
because each differentiation over $y$ in the left-hand side provides the
factor $q\gamma $ while the factors $K_q$ and $F_q$ appear in left- and right-hand
sides due to the differentiation over $u$.} as $q^{-2}$. Thus the role of genuine
correlations is strongly damped in systems with the large number of
particles.

However, the asymptotics is very far from our realm.
At present energies, according to QCD, this ratio should reveal the minimum
at $q\approx 5$ and subsequent oscillations. In gluodynamics, the minimum
is located at
\begin{equation}
q_{min}\approx \frac {24}{11\gamma _0}+0.5+O(\gamma _0).  \label{qmin}
\end{equation}

This astonishing qualitative
prediction \cite{13} of the new type of the moments behaviour has been
confirmed in experiment (for the very first time in Ref.
\cite{dabg}) as in $e^{+}e^{-}$ (see Fig. 5 \cite{sld}) as in hadronic
processes \cite{dnbs}. The predicted negative minimum of $H_q$ is clearly
observed. These oscillations can correspond to the replacement of attractive
forces (clustering) by repulsion (between clusters) in systems with different
number of particles. Let us note that the colour coherence in QCD discussed below
also predicts both the attractive and repulsive forces in the systems of 
the coloured partons.

With energy increasing, one should observe first the disappearence
of the oscillations at high ranks and then the smaller dip at the first
minimum and its slow shift to higher values of $q$ from its initial position
at $q\approx 5$ at Z$^0$ energy due to the decrease of $\gamma _0$
(see Eqn (\ref{qmin})). Finally, the ratio $H_q$ will slowly tend to
its asymptotic dependence $q^{-2}$.

The minimum position slowly changes with energy and with the size of the phase
space window because it is inverse proportional (see \cite{13, imdr}) to the
square root of the running coupling strength, i.e., to $\gamma _0$. For some
specific processes this shift can be very strong (e.g., it has been found for
instanton induced processes \cite{kshu} that the minimum moves to $q\approx 2$
because the multiplicity distribution in these processes is very narrow).
At the same time, other experimental data confirm that the oscillations
reveal themselves on the qualitative level in various processes, i.e., have 
some universal origin \cite{dnbs}. The graphical presentation of the ratio
$H_q$ in place of the moments is suitable, because the moments increase fast
with the rank $q$ increasing but this growth cancels in their ratio.

The quantitative analytical estimates are not enough
accurate because, first of all, the expansion parameter becomes equal to the
product $q\gamma $ which is close to one or even exceeds it for all $q>1$.
Therefore the perturbative approach is, strictly speaking, inapplicable to
this problem. However, some tricks like Pade-approximation can be used to
improve it. At the same time,
the numerical computer solution \cite{lo2, lo1} reproduces oscillations quite well.

These new dependences differ from all those typical for the previously attempted
distributions of the probability theory, in particular, such previously popular
in particle physics distributions as the Poisson and negative binomial ones.
The characteristics which are special for the newly found distribution at high
energies have been discussed in Ref. \cite{bkta}.

\subsection{The hump-backed plateau}

Dealing with inclusive distributions of partons within the jet over the
rapidity or
over the share of the jet energy ($x$) acquired by the parton, one should solve
equations of the type (\ref{symb}) for the
generating functional. The variation over the auxiliary function $u(k)$ gives
rise to integro-differential equations for the one-particle distributions
and the correlation functions. In the leading order in gluodynamics the equation
for the distribution function $D(x,y)$ is written as
\begin{equation}
\frac {d(xD(x,y))}{dy}=\frac {1}{4N_c}\int _0^1dz\gamma _0^2K(z)\left [\frac {x}{z}
D(\frac {x}{z},y+\ln z)\right ]
\end{equation}
with the boundary condition $xD(x,0)=\delta (1-x)$. Here, $x=k/E_j$,
$k$ is the particle momentum, $E_j$ is the jet energy, $K(z)=K^G_G(z)$.
In QCD there are two equations for gluon and quark jets. The subsequent use
of the Mellin transformation allows one to reduce these equations in the low
 orders to the solvable differential equations.
This has been done up to the NLO approximation (see \cite{5, kowo}).

As predicted by QCD, the momentum (or, more accurately, rapidity $y_c$\footnote{
By the index $c$, we denote the values of the rapidity in the center of mass
system.}) spectra
of the particles inside jets in $e^{+}e^{-}$-annihilation processes should have
the shape of the hump-backed plateau \cite{adk1, dfkh, bcmm, adkt}.
This striking prediction of the perturbative QCD differs from the previously
popular flat plateau advocated by Feynman.
This has been found in experiment (Fig. 6).
The depletion between the two humps is due to the angular ordering and colour
coherence in QCD. The humps are of the approximately Gaussian shape
(the distorted Gaussian) near
their maxima if the variable $\xi=\ln \frac {1}{x}$
is used. This
prediction was first obtained in the LO QCD, and more accurate expressions were
derived in NLO \cite{fweb}.

The maximum position $\xi _0$ in the $\xi $-distribution must shift with energy
increase according to the almost logarithmical law with some corrections of the 
type $O(\ln ^{-1/2}s)$:
\begin{equation}
\xi _0=y\left [\frac {1}{2}+\sqrt \frac {C}{y}- \frac {C}{y}\right ],
\end{equation}
where $C=\frac {a^2}{16N_c\beta _0}(\approx 0.3); \; a=\frac {11}{3}N_c+\frac
{2n_f}{3N_c^2}$.

Moments of the distributions up to the fourth rank have been calculated. The
spectrum drops quite fast towards smaller momenta. This is a consequence of
the colour coherence for the ever softer gluons. It becomes especially noticeable
if the variable $\xi $ is used. In the absence of the colour coherence the shape
of the distribution near the maxima losts its Gaussian form, the maxima are
shifted to the region of smaller values of $x$ (larger $\xi $) and, therefore,
the spectrum becomes much softer.

The comparison of these QCD predictions with experimental data at different
energies has revealed good agreement both on the shape of the spectrum (see
Fig. 7 for $e^{+}e^{-}$ from \cite{delp}) and on the energy dependences of its
peak position (see Fig. 8 for $e^{+}e^{-}, ep, p\overline p$
from \cite{cdf}) and of its width.

For soft particles, the spectrum does not depend on the energy \cite{adkt} and
has the shape \cite{loc1, loc2, klo1, klo2}:
\begin{equation}
\frac {dn}{dy_cd\ln k_t}\propto \alpha _S(k_t).
\end{equation}
This explains the discussed above possibility to observe the approach to the
asymptotic value of the ratio $r$  even at finite energies if the soft particles
with $p_t\rightarrow Q_0$ are chosen. The physical origin of this effect lies
in the ability of soft partons to resolve only the total charge of the jet
(i.e., of the parton initiating it) but not its internal structure.

\subsection{Difference between heavy- and light-quark jets}

Another spectacular prediction of QCD is the difference between the spectra
and multiplicities in jets initiated by heavy and light quarks. Qualitatively,
this corresponds to the difference in the bremsstrahlung by muons and electrons 
where the photon emission at small angles is strongly suppressed for muons
because of the large mass in the muon propagator. Therefore, the intensity of
the radiation is lower in the proportion to the ratio of the masses squared.
The suppression of
the photon radiation at small angles was proposed to be exploited \cite{dnaz}
also for the top quark where it is especially strong due to its large mass.
However, the intensity of the radiation would be too low due to the 
same reason.

The coherence of soft
gluons also plays an important role in QCD. For heavy quarks the accompanying
radiation of gluons should be stronger depleted in the forward direction
(the so-called dead-cone or ring-like emission). At large angles there should
be no visible difference between the emission of gluons by heavy or light
quarks. In spite of the close analogy, there is a strong difference between 
quantum chromodynamics and quantum electrodynamics connected to the fact that
the gluons possess the self-interaction while the photons have no point-like
interaction of one with another.

It was predicted \cite{83, sdkk, 87} that the coherence of the soft gluons
should result in the energy-independent difference of companion mean
multiplicities for heavy- and light-quark jets of equal energy. The companion
hadron multiplicity for the heavy quark is defined as the difference between
the total multiplicity of the jet initiated by this quark and the constant
decay multiplicity $n^{dc}$ in its rest system (which is equal to
11.0$\pm $0.2 for the $b$-quark and 5.2$\pm $0.3 for the $c$-quark).
Correspondingly, the difference of the total average multiplicities $n$ is 
written as
\begin{equation}
n_{Q\bar Q}(E)-n_{q\bar q}(E)=n^{dc}_{Q\bar Q}(M)-n_{q\bar q}(\sqrt eM)[1+O(\alpha _S(M))].
\end{equation}
Here, $q, Q$ denote the light and heavy quarks, $E$ are their energies, $M$ is 
the heavy quark mass, $\sqrt e\approx 1.65$. The naive model of energy
rescaling \cite{85, 86} predicts
the decreasing difference at higher energies. The experimental data (see Fig. 9
from \cite{del1}) comparing the multiplicities in jets initiated by light
($u, d, s$) and heavy ($b, c$) quarks support this QCD conclusion.

Another consequence of such a distinction is the effect of the leading heavy
quarks, i.e., the consumption by them of the rather large share of the primary
momentum in the process of the emission of gluons. This is related to the
suppression of the emission of hard gluons with low transverse momenta which
results in the low energy loss by heavy quarks. This effect has also been
observed \cite{delh} in $e^{+}e^{-}$-annihilation processes. The inclusive
spectra of the $b$-quark jet at the energy of the $Z^0$ have the maxima at
$x_Q\approx 0.8 - 0.9$. In the hadronic processes, this effect was
earlier used \cite{dyak} to explain the so-called "long-flying cascades" in 
cosmic rays at the energy about 100 TeV. In result, the conclusion about
the strong increase of the total cross section for the production of heavy
quarks in the energy interval up to 100 TeV was obtained. It finds now
the direct support from accelerator data.

Also, it has been noticed \cite{opa2} that the momentum spectra of particles in
the $b$-jets are much softer than those in the jets initiated by light quarks
as has been predicted in QCD \cite{83} with account of the coherence.

Concerning the difference of the angular distributions for gluon emission by
heavy and light quarks initiating the cascade, which lies at the background
of all these effects, there exist just the preliminary results \cite{nomo, del1},
which favour the QCD predictions.

Let us note that for hadronic processes the ring-like structure of the polar
angle distribution can be observed also for the jets themselves, i.e., for
the partons initiating these jets. It can result due to the limitations 
imposed on the radiation length or due to the so-called Cherenkov gluon effect
\cite{dr79, dikk} what corresponds in the "operationalists" language to the
emergence of some effective mass.

\subsection{Colour coherence in 3-jet events}

When three or more partons are involved in a hard interaction, one should take 
into account colour-coherence effects. They depend on the event topology and,
in particular, become stronger for smaller angles between the two jets
\cite{139}. For example, for two jets moving quite close to each other, one
should consider their mutual screening. It implies that for the "resolution"
of such a pair one should use the rather "hard probe" because the soft ones
will feel only the total colour charge of the pair as a whole. In particular,
it results in the "colour transparency" of such a pair during its motion inside
a hadronic medium. In electrodynamics, the similar effect of the mutual
screening of the electric charges of the closely moving electron and positron
until they separate at a large distance is known as Chudakov effect \cite{47}.

In equation for the generating functionals one should take into account 
those correlations between the jets which arise if the Feynman diagrams are
considered. However, the probabilistic interpretation in the framework of
equations (\ref{50}), (\ref{51}) in some approximation may fail for these
effects. Therefore, for studies of the colour coherence effects the Feynman
diagram technique is preferred.

Several such effects have been already observed. In particular, the total
multiplicity can not be represented simply as a sum of flows from independent
partons. In terms of the observed multiplicities $n_{e^{+}e^{-}}$ in
annihilation processes, one can write down the multiplicity in the three-jet
events $n_{q\bar qg}$ \cite{koch} as
\begin{equation}
n_{q\bar qg}=[n_{e^{+}e^{-}}(2E_q)+0.5r(p_{t})n_{e^{+}e^{-}}(p_{t})](1+
O(\alpha _S)), \label{nqqg}
\end{equation}
where $E_q$ is the quark energy and $p_t$ is the transverse momentum of the
gluon in the center of mass system for the pair $q\bar q$. The ratio of the 
multiplicities for gluon and quark jets is considered at the energy equal to the
transverse momentum of the gluon which, in its turn, is related to the
virtuality of the quark emitting this gluon. The existence of two scales
in the three-jet events is clearly demonstrated in this way. The comparison
of formula (\ref{nqqg}) with experimental data \cite{115, 188} has shown
that it is valid for large angles between the jets.

QCD predicts
that the particle flows should be enlarged in the directions of emission of
partons and suppressed in between them. Especially interesting is the prediction 
that due to the negative interference this suppression is stronger between
the $q\overline q$-pair than between $gq$ and $g\overline q$ in the hard
$e^{+}e^{-}\rightarrow q\overline {q}g$ event if all angles between partons are
large. This phenomenon is known as the "string" \cite{agsj} or "drag" \cite{64}
effect. All these predictions have been confirmed by experiment (see Fig. 10
from \cite{del2}). In $q\overline qg$ events the particle population values
in the $qg$ valleys are found larger than in the $q\overline q$ valley by a
factor 2.23$\pm $0.37 compared to the theoretical prediction of 2.4. Moreover,
QCD predicts that this shape is energy-independent up to an overall
normalization factor.

Let us note that for the process $e^{+}e^{-}\rightarrow q\overline q\gamma $
the emission of additional photons would be suppressed both in the direction of
a primary photon and in the opposite one. In contrast, in the case of the
emitted gluon one
observes the string (drag) effect of enlarged multiplicity in its direction and
stronger suppression in the opposite one. This suppression is described by the
ratio of the corresponding multiplicities in the $q\overline q$ region
\begin{equation}
R_{\gamma }=\frac {N_{q\overline q}(q\overline qg)}
{N_{q\overline q}(q\overline q\gamma )}
\end{equation}
which is found to be equal 0.58$\pm $0.06 in experiment whereas the theoretical
prediction is 0.61.

The colour coherence reveals itself as inside jets as in the inter-jet regions.
It should suppress both the total multiplicity of
$q\overline {q}g$ events and the particle yield in the transverse to the
$q\overline {q}g$ plane for decreasing opening angle between the low-energy
jets. When the hard gluon becomes softer, colour coherence determines, e.g., the
azimuthal correlations of two gluons in $q\overline qgg$ system. In particular,
back-to-back configuration ($\varphi \sim 180^0$) is suppressed by a factor
$\sim 0.785$ in experiment, 0.8 in HERWIG Monte Carlo and 0.93 in analytical
pQCD.

Let us stress once again that the colour coherence determines the topological
dependence of the jet properties predefined theoretically in terms of the parton
diagrams. The interference between the $q\bar q$-pairs not connected by colour
is suppressed by the factor $1/N_c^2$. This interference is not accounted by
the models using the Monte Carlo method. Thus their success in the description
of experimental data implies the smallness of such effects. Nevertheless, 
these small colour-suppressed effects disappearing at $N_c\rightarrow \infty $
can become really important for the distinction between the analytic
diagram approach in QCD and the purely probabilistic Monte Carlo schemes
\cite{155}. The study of the individual events with very high multiplicity 
may be crucial for getting the decisive conclusions. The most important lesson
derived from the correspondence between the theory and experiment in this
case consists in the conclusion that the colour coherence leads to the effects
observed at the hadron level and is not wiped out by the hadronization stage.
The hadron distributions depend on the topology of the parton stage with hard 
colour objects that gives further support to the local parton-hadron duality 
hypothesis.

Some proposals have been promoted for the modification of formula (\ref{nqqg})
with more correct account for the phase space and for
special two-scale (the energy and the transverse momentum) analysis of 3-jet
events when the restriction on the transverse momentum of a gluon jet is 
imposed \cite{139, egus, egkh}. This corresponds to the simultaneous account for
the energy and the virtuality of the initially produced quark and antiquark.
The first approbation of this proposal \cite{171} has shown the correctness
of such a modification. This is related to the solution of the 
problem of the unique separation of the three-jet events according
to the kinematics at the parton level and clear detection of the gluon jet in
this system mentioned above. In general, jets are biased because event selection
according to some algorithms introduces bias
on multiplicities. Jet properties depend on two scales, the transverse momentum
and available rapidity range. To compare with theory, one should get unbiased
results. For gluon jets, they have been obtained in \cite{488} using the
following formula
\begin{equation}
n_{gg}(p_t)=2[n_{q\bar qg}(s, p_t)-n_{q\bar q}(s, p_t)],
\end{equation}
where the gluon transverse momentum is given by
\begin{equation}
p_t^2s=s_{qg}s_{\bar qg}
\end{equation}
and $s_{ig}$ is the squared cms energy of the system $ig$. This multiplicity
depends on a single variable, whereas the terms in the right hand side depend
also on the energies. This implies that there is no dependence on the algorithm
of jets separation. It is this accounting of the dependence on the cutoff
of the transverse momentum, below which the gluon jet is not resolved and the
whole system is treated as a $q\bar q$-pair, that finally leads to the
independence of the multiplicity of the gluon jet on the adopted algorithm
of jet separation.

Let us note that the fast divergence of the modified perturbation series with
increase of the term order $n$, which is like $n!$, usually related to the
notion of renormalons \cite{bene}, gives rise to large non-perturbative
corrections. This divergence was noticed above as the appearance of $q\gamma $
factor as the expansion parameter for high ranks of the moments of the
distributions.

The non-perturbative corrections to event shapes (especially, to three-jet
events) should be mentioned here. Just in this case these corrections are
very strong, and there are numerous experimental data here. In distinction
to the logarithmic dependences of the perturbative approach, the
non-perturbative terms decrease in a power-like manner with increase of the
transferred momentum. They can be written in a universal way \cite{yura}
for various characteristics $\nu _i$ as
\begin{equation}
\delta \nu _i \propto c_i (\Lambda _{eff}(p)/Q^2)^p,  \label{nu}
\end{equation}
where the parameter $\Lambda _{eff}$ does not depend on a particular
characteristics of the event shape $\nu _i$, which can be  the jet mass,
thrust, jet widening etc. The numerical value of the parameter $c_i$ is
calculated in QCD \cite{yura}.

Even though such an approach is quite successful in more precise description
of experimental data, there are many unsolved problems. In particular,
it is not clear where one should cut off the perturbative series and add
to it the non-perturbative terms, what is the relative role of corrections with
increasing values of $p$ at lower transferred momenta $Q^2$ and where, in
general, the non-perturbative terms (\ref{nu}) should saturate and flatten out.
Moreover, it was supposed \cite{hama} that one should use the renormalization
group improved perturbative expansion and it was shown that this improvement
leads to very satisfactory description of experimental data.

Results for jets in $ep$ and $p\overline p$ processes also favour the
theoretical expectations of the role of coherence effects for emission of soft
gluons. Here, analysis is complicated by the internal structure of the
colliding objects. Especially interesting are the data \cite{162, 182} on the
topology of the $p\bar p$-events either with the production of the $W$-meson in 
combination with the quark or gluon jet or with the production of two hadronic
jets \cite{klo1, 180}.

\subsection{Intermittency and fractality}

The self-similar parton cascade leads to special multiparton correlations. Its
structure with "jets inside jets inside jets..." caused by the angular
ordering has provoked the analogy with
turbulence and the ideas of intermittency \cite{66} according to which the 
increase of fluctuations in ever smaller phase space volume studied (e.g., in
smaller rapidity intervals $\delta y$) must lead to the increase of factorial
moments according to the following power-like behaviour:
\begin{equation}
F_q\propto (\delta y)^{-\phi (q)}.      \label{phi}
\end{equation}

In its turn, this self-similar structure should result in a definite geometric
pattern. Namely, this leads to the fractal distribution
of particles inside their available phase space \cite{drje}. The notion of
fractality allows one to quantify the characteristics of the process
expressing them in terms of the fractal (or multifractal, Renyi) dimensions.
Especially important is the fact that the geometric dimensions are related
to the physical characteristics, namely, to the intermittency exponent
$\phi (q)$, which shows the slope of the increase of the factorial moments on
the doubly logarithmic scale (see Eqn (\ref{phi})).
The slopes
(the intermittency exponents) $\phi (q)$ for
different ranks $q$ are related to the Renyi dimensions $D_q$ in a following way:
\begin{equation}
\phi (q) = (q-1)(D-D_{q}) ,  \label{110}
\end{equation}
where $D$ is the topological dimension of the analyzed phase space windows 
(for example, $D$=1 if the dependence of the factorial moments on the length
of the rapidity interval is studied).

To calculate these characteristics, one uses the diagramatic approach in the
same way as it was described above when discussing the colour coherence effects.
This is necessary now because one has to deal with a small part of the total number
of the created partons within the fixed small phase space volume. The fractal
behaviour is usually defined by the dependence of the logarithm of the factorial
moments in function of the logarithm of the size of the chosen phase space
region.
The (mono)fractal behaviour would display the linear dependence of logarithms of
factorial moments on the logarithmic size of phase space windows. The moments
are larger in smaller windows, i.e. the fluctuations increase in smaller bins
in a selfsimilar power-like manner if the (mono)fractal distribution is studied
(see the review paper \cite{14}).

In QCD, the power dependence of moments on the size of the phase space
window appears for a fixed coupling regime \cite{21}.
In this case, the
monofractal behaviour with a constant Renyi dimension is pronounced:
\begin{equation}
D_{q} = \frac {q+1}{q} \gamma _0 
\end{equation}
and, correspondingly, the intermittency exponent $\phi (q) =$ const at any fixed
rank $q$.

The running coupling strength has a definite scale and, therefore, it
leads to some decline from this simple self-similar (monofractal) behaviour.
The running property of the coupling strength in QCD flattens off
\cite{38, 68, 70} this dependence at
smaller bins, i.e. the multifractal behaviour takes over there.

 Both the linear (in the doubly logarithmic scale)
increase at comparatively large but decreasing bins and its flattening for very
small bins have been observed in experiment (see Fig. 11 from \cite{opal}).
However, only
qualitative agreement with analytical predictions can be claimed here. The
higher order calculations are rather complicated and the results of LO with
some NLO corrections are yet mostly available. In experiment, the different cuts
have been used which hamper the direct comparison. However, those Monte Carlo
models where these cuts can be done agree with experiment. The relative role
of the partonic and hadronization stages of the cascade in this regime as well
as the applicability of the local parton hadron duality hypothesis to the
correlation characteristics of the process are still debatable.

The transparent interpretation of the observed effect in terms of the fractal 
phase space volume has been proposed in the framework of the Lund dipole
cascade model \cite{1121, 1122}.

\subsection{The energy behaviour of higher moments of multiplicity distributions}

Differentiating both sides of equations (\ref{50}), (\ref{51}) over $u$ and
using formulas (\ref{4}), (\ref{5}), one can get the equations for the
moments of multiplicity distributions of any rank. Their solutions would
describe the behaviour of the
multiplicity fluctuations. They tell us that the fluctuations of the
multiplicity of individual events must be larger for quark jets as compared to
gluon jets. Therefore, the moments of their distributions in quark jets are
larger than the corresponding moments for gluon jets. This tendency is clearly
seen in experimental data \cite{del2}.
The factorial moments increase both with their rank and with energy increasing.
In the perturbative expansion one gets the formulas \cite{dlne}, similar to 
formula (\ref{Y}) for $r$, and the energy increase of the moments is determined
by the decrease of $\gamma _0$. The corresponding coefficients in front of the 
terms $\gamma _0^n$ are calculated in \cite{dlne}.
From the mentioned above behaviour of $H_q$-moments one easily guesses that the
same is true for the cumulant moments with some difference.

The experimental results for the second rank factorial moments of 41.8 GeV 
gluon jets $F_2^G=1.023$ and for 45.6 GeV $uds$ quark jets, $F_2^F=1.082$ are
much smaller than the known long ago (see, e.g., \cite{1})
asymptotic predictions, viz. 1.33 and 1.75,
respectively. The NLO terms improve the description of the data compared to the
leading order results. If one accepts the effective value of $\alpha _S$ 
averaged over all the energies of the partons during the jet evolution to be
$\alpha _S\approx 0.2$, one obtains \cite{dlne} the NLO values $F_2^G\approx 1.039$ and
$F_2^F\approx 1.068$ at these energies which are quite close to the experimental
results. In this sense the NLO prediction can be said to describe the widths
of the gluon and quark jet multiplicity distributions at the $Z^0$ energy to
within 10$\%$ accuracy.

Unfortunately, the 2NLO and 3NLO terms worsen the agreement with data compared
to NLO (but not compared to LO) results \cite{dlne}.
The same is true for the higher order moments. It raises the general theoretical
problem of the convergence of the perturbative
expansion in view of the large expansion parameter $q\gamma $ mentioned above.
The attempts to account for conservation laws more accurately by
the modified evolution equations for high moments \cite{dede} have not led to
the success yet. It is remarkable that the computer solution of the QCD
equations \cite{lo2} provides a
near-perfect description of the higher moments as well. This suggests that
the failure of the analytical approach at higher orders is mainly a technical 
issue related to an inadequate treatment of soft gluons and of energy-momentum
conservation. This conclusion is also
supported by the success of the Monte Carlo model ARIADNE \cite{173} in
describing these characteristics.

The rather accurate experimental results about the multiplicity moments of the
separated gluon jets \cite{151} are available now up to the rank 5, and for
the quark jets \cite{sld} even up to the rank 17, because the accuracy of
measuring the multiplicity distributions for the latter is much higher due to
the larger statistics of the data.

The higher moments of the multiplicity distributions are determined by the
integrals of the correlation functions for partons inside the jet which depend
on their angular distributions. The angular correlations have been studied as
well \cite{111}. The comparison with experimental data \cite{117, 118} shows
that the agreement becomes better with the energy increase even though it is
still unsatisfactory at small angles. Some "infrared" stable characteristics
like correlations of the energy and multiplicity flows have been considered
(see, e.g., \cite{119, 120}). They do not require for the determination of
the jet axis, as in the case of the angular correlations, where the accuracy
of this determination is crucial for the correlation studies at small angles.

In the case of the fixed coupling constant, these equations are exactly
solvable \cite{21, dhwa}. Their solutions as functions of energy possess the
scaling property. All moments behave in a power-like manner with energy
increasing.

Let us make two short technical comments at the end of this subsection. All
analytical expressions should satisfy the requirements of the limiting case of
the supersymmetric QCD (the mote detailed discussion of this problem and the
corresponding formulas for the coefficients $a_i$ and $r_i$ from the Table~1
see in Ref. \cite{cdnt8}). The factorial moments are always positive
according to their definition. These two requirements are the necessary but
not sufficient conditions for the correctness of the calculations. However,
the moments calculated in Ref. \cite{mweb} do not satisfy the second condition.
At the same time, the coefficient $r_2$ obtained in Ref. \cite{gmue} satisfies
the requirements of the supersymmetric QCD (as well as that from Ref. 
\cite{cdnt8} which differs numerically) but the method of the renormalization
group used in Ref. \cite{gmue} does not take into account correctly the
energy-momentum conservation laws and gives rise to the numerically smaller
value of this coefficient.

\subsection{Subjet multiplicities}

A single quark-antiquark pair is initially created in $e^{+}e^{-}$-annihilation.
With very low angular resolution
(large angle averaging) one observes two jets. A three-jet structure
can be observed if a gluon with large transverse momentum is emitted by the
quark or antiquark. However such a process is suppressed by an additional
factor $\alpha _S$, which is small for large transferred momenta. Its
probability can be
calculated perturbatively. At relatively low transferred momenta, the jet
evolves to the angular ordered subjets ("jets inside jets inside jets...").
Different algorithms have been proposed to resolve subjets. By increasing the
resolution, more and more subjets are observed. For very high resolution, the
final hadrons are resolved. The resolution criteria are chosen to provide the
infrared safe results (see, e.g., the papers \cite{231, 201, 211, 221, 261, 271}).

In particular, one can predict the asymptotic ratio of subjet multiplicities
in 3- and 2-jet events if one neglects the soft gluon coherence:
\begin{equation}
\frac {n_3^{sj}}{n_2^{sj}}=\frac {2C_F+C_A}{2C_F}=\frac {17}{8}.
\end{equation}
Actually, the coherence reduces this value to be below 1.5 in experiment for
all acceptable resolution parameters. The theoretical predictions \cite{cdfw}
agree quantitatively with experimental findings \cite{l3, opa1} at the lowest
resolution, where this ratio is equal to 3/2, and only qualitatively at higher
resolutions. The computer calculations \cite{lo2, lo1} agree quite well with 
experiment for different resolution parameters that implies the importance 
of the precise account for the energy-momentum conservation.

Subjet multiplicities have also been studied for the separated quark and gluon jets.
The analytical results \cite{seym} are seen (Fig. 12 from \cite{alep}) to
represent the data fairly well for
large values of the subjet resolution scale $y_0$.

\subsection{Jet universality}

According to QCD, jets produced in processes initiated by different colliding
particles must be universal and depend only on their own parent (gluon, light
or heavy quark). This prediction has been confirmed by many experiments.
For example, this is clearly seen if the multiplicities in the fragmentation
region in the Breit system of $ep$-collisions are compared with the corresponding
multiplicities in $e^{+}e^{-}$-processes at high energies. It has been found
\cite{085} that they coincide as is required by the universality condition.
In this review paper, e.g., this jet universality was already mentioned in
several subsections and demonstrated in Fig. 8. Therefore we will not discuss
it at some length but just stress once again the importance of such a prediction
and its non-trivial origin. At the same time, let us mention that the 
universality of the properties of the jets produced does not imply the
universality of the mechanisms of their production in various processes.

\section{Conclusions and outlook}
                                                  
A list of successful analytical QCD predictions can be made longer. As was
demonstrated above, quantum chromodynamics has already predicted spectacular
qualitative features of rather soft processes. Quantitatively, analytical
results show that the higher order (NLO) terms always tend to improve the
agreement with experiment compared to the asymptotic
(LO) predictions. The accuracy achieved is often better than 20$\%$ or even
10$\%$ that is surprising by itself considering the rather large values of the
expansion parameter
of the perturbative approach. Moreover, some characteristics are very sensitive
to ever higher order terms and should be carefully studied. The astonishing
success of the computer solutions and Monte Carlo schemes demonstrates the
importance of using the correct borderline between the perturbative and
non-perturbative regions, which is approximately accounted in the analytical
perturbative approach by the cut-off parameter $Q_0$ and by the
limits of integration over parton splitting variables. One can expect that
the purely perturbative description becomes dual to the sum over all possible
hadronic excitations. Nevertheless, the correspondence between the parton stage
of the cascade evolution and the hadronization is not always defined at the
quantitative level and requires further studies and the development of the
common point of view. In particular, this problem is approached in a different
way not only in analytical results if they are compared with the Monte Carlo
models' conclusions but also in different Monte Carlo schemes.\\

A new era of multiparticle production studies opens with the advent of new
accelerators RHIC, LHC, TESLA, NLC, JLC, CLIC. We come closer to the asymptotic
region\footnote{Let allow me to make here some short historical and lyrical
digression. About 45 years ago, the tuitor of my diploma work I.Ya. Pomeranchuk
told me that with the advent of the 10 GeV Dubna accelerator we enter the
asymptotic region because $10\gg 1$ (By 1 he meant the nucleon mass.).
Now, we understand that this hope was too naive. Probably, our today's
feelings are not more satisfactory even though they are supported by the
modern QCD predictions. However, possible effects of the high parton density
and of the non-linear interactions, which nowadays escape the detailed
treatment, could change the situation at high energies.} even though the
approach to the asymptotic laws will be, probably, extremely slow because
all the predicted energy dependences of the physically measurable quantities
are slow as well. Nevertheless, some predictions differ for various analytical
approaches and Monte Carlo schemes at these energies and will be confronted
to experimental data. It will allow to distinguish between them.
The qualitative QCD predictions indicate the tendencies towards the asymptotic
region where the perturbative estimates become more precise.

The mean multiplicities will increase drastically. Now, in Au-Au collisions
at the center of mass energies of 130 GeV per nucleon at RHIC the mean charged
multiplicity exceeds 4000.
It implies that the event-by-event analysis of various patterns formed by
particles in the available phase space becomes meaningful. Such an analysis
would allow one to classify in more detail the multiparticle production 
processes than, say, it is done in hadronic interactions where simple
separation of the diffractive and non-diffractive processes is only considered.
The study of the topology of individual events must provide us with a much
richer information on the dynamics of the process compared with the results
of the measurement of characteristics averaged over the whole event sample.
Its results can be compared to the exclusive probabilistic Monte Carlo schemes.
Searches for the supersymmetric partners of well known particles, Higgses,
new states of matter, new collective and interference effects,
physics of jets and mini-jets will, surely, be among the most important
directions of further investigations.

The event-by-event approach would allow one to analyze the small
colour-suppressed effects, which show the difference between the perturbative
QCD and Monte Carlo calculations for the topology of the individual events,
properties of minijets or clusters (with the attraction-repulsion transition),
other collective effects like the elliptic flow (and even the higher Fourier
expansion terms of the azimuthal distribution in an individual event),
the possible azimuthal asymmetry of the opening angle for
individual jets, the ring-like events (the probable signature of the
confinement and/or of the "Cherenkov gluons"), further analysis of the
difference in accompanying gluon emission by heavy and light quarks (in
particular, the angular distribution difference) etc.

Of the principal importance is the study of the multiparticle correlations.
The use of traditional formulas for the correlation functions is hampered by
the large number of the independent variables. One can overcome this difficulty
in spite of the seemingly complicated structure of these functions.
The event-by-event analysis
of experimental exclusive data can become available, quantified locally and
provide statistically significant results if one uses wavelets 
(sometimes called the "mathematical microscope") for the pattern recognition in
individual events \cite{dine}. With the help of wavelets one can separate
correlations at the different resolution levels (from the short-range to
long-range correlations) locally and in a compact form. In principle, the
distributions of the wavelet coefficients can replace the complicated
expressions for the multiparticle correlators.

At the same time, the intriguing data obtained in the cosmic ray studies
(for the recent review see Ref. \cite{glad})
tell us that one can await for new phenomena in the fragmentation region
of hadronic (and nuclear) collisions which, unfortunately, has not been
devoted sufficient attention in accelerator experiments because most detectors
are suited for the central range of rapidities. In particular, it would be
important to learn the ratio between the baryon, meson and photon contents
of the multiparticle production events.

To confront QCD predictions with new
experimental findings at ever higher energies will be crucial for the
search for possible new states of matter. This is also important for the
separation of any new physical signals from the conventional QCD background.\\

{\bf Acknowledgements}\\

I am deeply indebted to all colleagues with whom I worked on the problems
reviewed in this paper, and to the authors of those numerous papers which often
stimulated this research. This work was supported by the RFBR grant 00-02-16101.


{\bf Figure captions}\\

Fig. 1. The energy dependence of average multiplicity  of charged particles in 
$e^+e^-$-annihilation. The results of different fits according to formulas of
perturbative QCD and of the Monte Carlo models are shown ( the solid and dotted
lines are the fits of formula (\ref{mean}) with one and two adjusted parameters,
the dashed line is given by the HERWIG Monte Carlo model; the vertically shaded
area indicates the gluon jet data multiplied by the theoretical value of the
ratio $r$ (\ref{Y})).\\

Fig. 2. The experimentally measured ratio $r$ of multiplicities in gluon and quark
jets as a function of energy in comparison with the predictions of analytical QCD
and of the Monte Carlo model HERWIG (different QCD approximations, described
in this paper, as well as 3NLO($\epsilon $) with integration limits
$e^{-y}$ and 1-$e^{-y}$ in Eqns (\ref{50}), (\ref{51}) are indicated at
the corresponding lines).\\

Fig. 3. The ratio of the slopes of the energy dependences of mean
multiplicities in gluon and quark jets according to experimental data and some
theoretical calculations.\\

Fig. 4. The ratio of the curvatures of the energy dependences of mean
multiplicities in gluon and quark jets according to experimental data and some
theoretical calculations.\\

Fig. 5. The measured ratio $H_q$ of the cumulant and factorial moments oscillates
as a function of the rank $q$ according to experimental data on multiplicity
distributions of charged particles in $e^+e^-$-annihilation at the $Z^0$ energy
(the inset in the upper right corner shows the data for the moments of
the ranks 2, 3 and 4).\\

Fig. 6. The inclusive rapidity distribution of secondary particles in
$e^+e^-$-annihilation  has the shape of the hump-backed plateau (it is shown
in the center of mass system).\\

Fig. 7. The peak in the variable $\xi $ is fitted at different energies by the
distorted Gaussian with the moments predicted by the NLO-approximation of
perturbative QCD (the fitted lines are drawn at $\Lambda =210$ Β') according
to \cite{fweb}).\\

Fig. 8. The peak position of the experimental inclusive distribution as a function
of the mass of two jets is compared with NLO predictions (the central line
is fitted by the CDF Collaboration data only). The lowest order (LO) predictions
are shown by the lower straight line. The expected behaviour without the colour
coherence effect gives rise to the upper line.\\

Fig. 9. The energy dependence of the difference between the average multiplicities
of charged particles in events initiated by $b$ and ($u, d, s$)-quarks.
Experimental dots are compared with QCD predictions (the horizontal stripe)
and with the results of the naive rescaling model (the decreasing stripe).\\

Fig. 10. The charged hadronic flows in 3-jet events (the histogram) in comparison
with analytical QCD predictions (the solid line) as functions of the azimuthal
angle.\\

Fig. 11. The normalized factorial moments of various ranks as functions of their
scale (the size of the phase space window diminishes to the right on the
abscissa axis). Different analytical approximations are compared with
experimental data. LO-approximation (DLLA): (a) - \cite{38},
(b) - \cite{70}, (c) - \cite{68}; NLO-approximation (MLLA) - \cite{38}. The 
qualitative but not quantitaive agreement is seen.\\

Fig. 12. The subjet multiplicities in the separated gluon (a) and quark (b)
jets as functions of the resolution parameter $y_0$ are compared with
analytical QCD results and with predictions of the Monte Carlo model JETSET
for hadrons (HL) and partons (PL).\\


\begin{thebibliography}{99}
\bibitem{1}
Andreev I V {\it Chromodynamica i zhestkie processy pri vysokih energiyah}
(Moscow, Nauka, 1981) (in Russian)
\bibitem{2}
Ioffe B L, Lipatov L N and Khoze V A {\it Glubokoneuprugie processy} (Moscow,
Energoatomizdat, 1983)  (in Russian)
\bibitem{3}
Yndurain F J {\it Quantum chromodynamics} (N.-Y.-Berlin-Heidelberg-Tokyo,
Springer Verlag, 1983).
\bibitem{4}
Voloshin M B and Ter-Martirosyan K A {\it Teoriya kalibrovochnyh vzaimodeistvii
elementarnyh chastits} (Moscow, Energoatomizdat, 1984)  (in Russian)
\bibitem{5}
Dokshitzer Yu L, Khoze V A, Mueller A H and Troyan S I {\it Basics of
perturbative QCD} ed. J. Tran Thanh Van
(Gif-sur-Yvette, Editions Frontieres, 1991).
\bibitem{dre1}
Dremin I M  {\it UFN} {\bf 164} 785 (1994); {\it Phys.-Uspekhi} {\bf 37} 715 (1994)
\bibitem{koch}
Khoze V A and Ochs W {\it Int. J. Mod. Phys.} {\bf A 12} 2949 (1997) 
\bibitem{dgar}
Dremin I M and Gary J W {\it Phys. Rep.} {\bf 349} 301 (2001)
\bibitem{kowo}
Khoze V A, Ochs W and Wosiek J in {\it "Handbook of QCD" (Ioffe Festschrift)}
(WSPC, Singapore) (to be published); hep-ph/0009298
\bibitem{stir}
Stirling W J {\it J. Phys.} {\bf G 26} 471 (2000)
\bibitem{glry}
Gribov L V, Levin E M and Ryskin M G {\it Phys. Rep.} {\bf 100} 1 (1983)
\bibitem{lrys}
Levin E M and Ryskin M G {\it Phys. Rep.} {\bf 189} 267 (1990)
\bibitem{lipa}
Lipatov L N {\it Phys. Rep.} {\bf 286} 131 (1997)
\bibitem{fstr}
Frankfurt L L and Strikman M I {\it Phys. Rep.} {\bf 160} 235 (1988)
\bibitem{wang}
Wang X-N {\it Phys. Rep.} {\bf 280} 287 (1997)
\bibitem{schm}
Schmidt C R {\it Proc. RADCOR 2000}, Santa Cruz, CA, USA, 11-16 Sept 2000;
hep-ph/0106181
\bibitem{agis}
Andersson B, Gustafson G, Ingelman G and Sj\"{o}strand T {\it Phys. Rep.}
{\bf 97} 31 (1983)
\bibitem{mawe}
Marchesini G and Webber B {\it Nucl. Phys.} {\bf B 238} 1 (1984); 
{\bf B 310} 461 (1988)
\bibitem{mwak}
Marchesini G, Webber B, Abbiendi G et al {\it Comp. Phys. Comm.} {\bf 43}
465 (1992)
\bibitem{adkt}
Azimov Ya I, Dokshitzer Yu L, Khoze V A and Troyan S I {\it Z. Phys.} {\bf C 27}
65 (1985); {\bf C 31} 213 (1986)
\bibitem{aven}
Amati D and Veneziano G {\it Phys. Lett.} {\bf B 83} 87 (1979)      
\bibitem{bcma}
Bassetto A, Ciafaloni M and Marchesini G {\it Phys. Lett.} {\bf B 83} 207 (1979)
\bibitem{mtve}
Marchesini G, Trentadue L and Veneziano G {\it Nucl. Phys.} {\bf B 181} 335 (1981)
\bibitem{kuve}
Konishi K, Ukawa A and Veneziano G {\it Nucl. Phys.} {\bf B 157} 45 (1979)
\bibitem{44}
Mueller A H {\it Nucl. Phys.} {\bf B 213} 85 (1983); Erratum {\bf B 241}
141 (1984)
\bibitem{glip}
Gribov V N and Lipatov L N {\it YaF} {\bf 15} 1218 (1972); {\it Sov. J. Nucl. Phys.}
{\bf 15} 781 (1972)
\bibitem{apar}
Altarelli G and Parisi G {\it Nucl. Phys.} {\bf B 126} 298 (1977)
\bibitem{dok1}
Dokshitzer Yu L {\it ZhETF} {\bf 73} 1216 (1977); {\it Sov. Phys. JETP} {\bf 73} 641
(1977)
\bibitem{lip1}
Lipatov L N {\it YaF} {\bf 23} 642 (1976); {\it Sov. J. Nucl. Phys.} {\bf 23} 338
(1976)
\bibitem{klfa}
Kuraev E A, Lipatov L N and Fadin V S {\it ZhETf} {\bf 72} 377 (1977);
{\it Sov. Phys. JETP} {\bf 72} 199 (1977)
\bibitem{blip}
Balitsky Ya Ya and Lipatov L N  {\it YaF} {\bf 28} 1597 (1978); {\it Sov. J. Nucl. Phys.}
{\bf 28} 822 (1978)
\bibitem{ccfm}
Ciafaloni M {\it Nucl. Phys.} {\bf B 296} 49 (1988);\\
Catani S, Fiorani F and Marchesini G {\it Phys. Lett.} {\bf B 234} 339 (1990);
{\it Nucl. Phys.} {\bf B 336} 18 (1990)
\bibitem{gust}
Gustafson G {\it Proc. 30 Int. Symp. on Multiparticle Dynamics}, Tihany,
Hungary, 9-15 Oct 2000, Eds T Csorgo, S Hegyi, W Kittel, World Scientific,
Singapore, 2001, p 42;\\
{\it Proc. 31 Int. Symp. on Multiparticle Dynamics}, Datong,
China, 1-7 Sept 2001 (to be published)
\bibitem{lln}
Lipatov L N {\it Nucl Phys Proc Suppl} {\bf 99A} 175 (2001)
\bibitem{dlip}
De Vega and Lipatov L N hep-ph/0107225
\bibitem{mlve}
McLerran L and Venugopalan R {\it Phys. Rev.} {\bf D 49} 3352 (1994)
\bibitem{jklw}
Jalilian-Marian J, Kovner A, Leonidov A and Weigert H {\it Nucl. Phys.}
{\bf B 504} 415 (1997)
\bibitem{leon}
Leonidov A {\it Proc. Conf. "Quantum Field Theory and Strings", Moscow, June 2000}
(to be published)
\bibitem {jmkw}
Jalilian-Marian J, Kovner A and Weigert H {\it Phys. Rev.} {\bf D 59} 014015 (1999)
\bibitem{13}
Dremin I M {\it Phys. Lett.} {\bf B 313} 209 (1993)   
\bibitem{21}
Dremin I M and Hwa R C {\it Phys. Rev.} {\bf D 49} 5805 (1994)
\bibitem{dhwa}
Dremin I M and Hwa R C {\it Phys. Lett.} {\bf B 324} 477 (1994)
\bibitem{shir}
Shirkov D V {\it TMF} {\bf 127} 409 (2001)
\bibitem{marc}
Marchesini G {\it Nucl. Phys. Proc. Suppl.} {\bf 71} 85 (1999)
\bibitem{dmwe}
Dokshitzer Yu L, Marchesini G and Webber B R {\it JHEP} {\bf 9907}:012 (1999)
\bibitem{bdmz}
Banfi A, Dokshitzer Yu L, Marchesini G and Zanderighi G {\it JHEP} {\bf 0007}:002
(2000); {\it Phys. Lett.} {\bf B508} 269 (2001); {\it JHEP} {\bf 0103}:007 (2001)
\bibitem{123}
Dremin I M {\it Pis'ma v ZhETF} {\bf 31} 215 (1980); {\it JETP Lett.} {\bf 31}
185 (1980)
\bibitem{124}
Dremin I M and Leonidov A V {\it YaF} {\bf 35} 288 (1982); {\it Sov. J. Nucl. Phys.}
{\bf 35} 247 (1982)
\bibitem{125}
Leonidov A V and Ostrovsky D M  {\it YaF} {\bf 60} 185 (1997); {\it Phys. Atom. Nucl.}
{\bf 60} 119 (1997)
\bibitem{127}
Ellis J and Geiger K {\it Phys. Rev.} {\bf D 52} 1500 (1995)
\bibitem{128}
Ellis J and Geiger K {\it Nucl. Phys.} {\bf A 590} 609c (1995)
\bibitem{eden}
Eden P {\it Proc. XXXIV Moriond conf. "QCD and strong interactions"} March 1999,
 ed. J. Tran Thanh Van (Editions Frontieres, Gif-sur-Yvette, 1999)
\bibitem{dede}
Dremin I M and Eden P {\it Proc. 31 Int. Symp. on Multiparticle Dynamics}
(Datong, China, Sept 2001) (to be published)
\bibitem{lyan}
Yang C N and Lee T D {\it Phys. Rev.} {\bf 87} 404 (1952)
\bibitem{ylee}
Lee T D and Yang C N {\it Phys. Rev.} {\bf 87} 410 (1952) 
\bibitem{80}
Particle Data Group, Barnett R M et al {\it Phys. Rev.} {\bf D 54} 1 (1996)
\bibitem{7}
CLEO Collaboration, Alam M S et al {\it Phys. Rev.} {\bf D 46} 4822 (1992);
{\it Phys. Rev.} {\bf D 56} 17 (1997)
\bibitem{139}
Dokshitzer Yu L, Troyan S I and Khoze V A {\it YaF} {\bf 47} 1010 (1988);
{\it Sov. J. Nucl. Phys.} {\bf 47} 881 (1988)
\bibitem{140}
Gary J W {\it Phys. Rev.} {\bf D 49} 4503 (1994)
\bibitem{9}
OPAL Collaboration, Alexander G et al {\it Phys. Lett.} {\bf B 388} 659 (1996)
\bibitem{egus}
Eden P and Gustafson G {\it JHEP} {\bf 9809} 015 (1998)
\bibitem{egkh}
Eden P, Gustafson G and Khoze V A {\it Eur. Phys. J.} {\bf C 11} 345 (1999)
\bibitem{488}
Gary J W in {\it Proc. 31 Int. Symp. on Multiparticle Dynamics} 1-7 Sept. 2001,
Datong, China, Eds. Liu L., Wu Y., World Scientific, Singapore, 2002
(to be published); \\
OPAL Collaboration Physics Note PN488, 2001 
\bibitem{231}
Brown N and Stirling W J {\it Phys. Lett.} {\bf B 252} 657 (1990);
{\it Z. Phys.} {\bf C 53} 629 (1992)
\bibitem{mu1}
Mueller A H {\it Phys. Lett.} {\bf B 104} 161 (1981)
\bibitem{dfkh}
Dokshitzer Yu L, Fadin V S and Khoze V A {\it Z. Phys.} {\bf C 15} 335 (1982);
{\bf C 18} 83 (1983) 
\bibitem{bcmm}
Bassetto A, Ciafaloni M, Marchesini G and Mueller A H {\it Nucl. Phys.} {\bf B 207}
189 (1982)
\bibitem{web1}
Webber B R {\it Phys. Lett.} {\bf B 143} 501 (1984)
\bibitem{dktr}
Dokshitzer Yu L, Khoze V A and Troyan S I {\it Int. J. Mod. Phys.} {\bf A 7} 1875 (1992) 
\bibitem{cdfw}
Catani S, Dokshitzer Yu L, Fiorani F and Webber B R {\it Nucl. Phys.} {\bf B 377}
445 (1992)
\bibitem{dg}
Dremin I M and Gary J W {\it Phys. Lett.} {\bf B 459} 341 (1999) 
\bibitem{cdnt8}
Capella A, Dremin I M, Gary J W, Nechitailo V A and Tran Thanh Van J
{\it Phys. Rev.} {\bf D 61} 074009 (2000) 
\bibitem{lo2}
Lupia S {\it Phys. Lett.} {\bf B 439} 150 (1998);
{\it Proc. XXXIII Moriond conf. "QCD and strong interactions"}
March 1998, ed. J. Tran Thanh Van (Editions Frontieres, Gif-sur-Yvette, 1998)
p. 363
\bibitem{lo1}
Lupia S and Ochs W {\it Phys. Lett.} {\bf B 418} 214 (1998); {\it Nucl. Phys.
(Proc. Suppl.)} {\bf B 64} 74 (1998)
\bibitem{brgu}
Brodsky S J and Gunion J F {\it Phys. Rev. Lett.} {\bf 37} 402 (1976) 
\bibitem{43}
Mueller A H {\it Nucl. Phys.} {\bf B 241} 141 (1984) 
\bibitem{gmue}
Gaffney J B and Mueller A H {\it Nucl. Phys.} {\bf B 250} 109 (1985)
\bibitem{mweb}
Malaza E D and Webber B R {\it Nucl. Phys.} {\bf B 267} 702 (1986)
\bibitem{41}
Dremin I M and Nechitailo V A {\it Mod. Phys. Lett.} {\bf A 9} 1471 (1994);
{\it JETP Lett.} {\bf 58} 881 (1993)
\bibitem{drsl}
Dremin I M {\it Pis'ma v ZhETF} {\bf 68} 635 (1998); {\it JETP Lett.} {\bf 68} 559
(1998)
\bibitem{desl}
DELPHI Collaboration, Abreu P et al {\it Phys. Lett.} {\bf B 449} 383 (1999)
\bibitem{opsl}
OPAL Collaboration, Abbiendi G et al {\it CERN-EP-2000-070} 
\bibitem{klo1}
Khoze V A, Lupia S and Ochs W {\it Eur. Phys. J.} {\bf C 5} 77 (1998)
\bibitem{klo2}
Khoze V A, Lupia S and Ochs W {\it Phys. Lett.} {\bf B 394} 179 (1997)
\bibitem{soft}
OPAL Collaboration, Abbiendi G {\it Eur. Phys. J.} {\bf C 11} 217 (1999) 
\bibitem{dabg}
Dremin I M, Arena V, Boca G et al {\it Phys. Lett.} {\bf B 336} 119 (1994) 
\bibitem{sld}
SLD Collaboration, Abe K et al {\it Phys. Lett.} {\bf B 371} 149 (1996)
\bibitem{dnbs}
Dremin I M, Nechitailo V A, Biyajima M and Suzuki N {\it Phys. Lett.}
{\bf B 403} 149 (1997)
\bibitem{imdr}
Dremin I M {\it Phys. Lett.} {\bf B 341} 95 (1994)
\bibitem{kshu}
Kuvshinov V I and Shulyakovsky R G {\it Acta Phys. Pol.} {\bf B 30} 69 (1999)
\bibitem{bkta}
Brooks T C, Kowalski K L and Taylor C C {\it Phys. Rev.} {\bf D 56} 5857 (1997)
\bibitem{adk1}
Azimov Ya I, Dokshitzer Yu L and Khoze V A {\it Pis'ma v ZhETF} {\bf 35} 390 (1982);
{\it JETP Lett.} {\bf 35} 482 (1982)
\bibitem{fweb}
Fong C P and Webber B R {\it Phys. Lett.} {\bf B 229} 289 (1989); {\bf B 241} 255 (1990);
{\it Nucl. Phys.} {\bf B 355} 54 (1991)
\bibitem{delp}
DELPHI Collaboration, Abreu P et al {\it Phys. Lett.} {\bf B 459} 397 (1999)
\bibitem{cdf}
CDF Collaboration, Safonov A N {\it Nucl. Phys. (Proc. Suppl.)} {\bf B 86}
55 (2000) 
\bibitem{loc1}
Lupia S and Ochs W {\it Phys. Lett.} {\bf B 365} 339 (1996)
\bibitem{loc2}
Lupia S and Ochs W {\it Eur. Phys. J.} {\bf C 2} 307 (1998)
\bibitem{dnaz}
Dremin I M, Nazirov M T and Saakian V A {\it YaF} {\bf 42} 1010 (1985); 
{\it Sov. J. Nucl. Phys.} {\bf 42} 845 (1985)
\bibitem{83}
Dokshitzer Yu L, Khoze V A and Troyan S I {\it J. Phys.} {\bf G 17} 1481, 1602 (1991)
\bibitem{sdkk}
Schumm B A, Dokshitzer Yu L, Khoze V A and Koetke D S {\it Phys. Rev. Lett.}
{\bf 69} 3025 (1992) 
\bibitem{87}
Petrov V A and Kisselev A V {\it Z. Phys.} {\bf C 66} 453 (1995)
\bibitem{85}
Azimov Ya I, Dokshitzer Yu L and Khoze V A {\it Ÿ"} {\bf 36} 1510 (1982);
{\it Sov. J. Nucl. Phys.} {\bf 36} 878 (1982)
\bibitem{86}
Kisselev A V, Petrov V A and Yuschenko O P {\it Z. Phys.} {\bf C 41} 521 (1988)
\bibitem{del1}
DELPHI Collaboration, Abreu P et al {\it Phys. Lett.} {\bf B 479} 118 (2000);
Erratum - ibid {\bf B 492} 398 (2000)
\bibitem{delh}
DELPHI Collaboration, Abreu P et al {\it Z. Phys.} {\bf C 57} 181 (1993);\\
ALEPH Collaboration, Busculic D et al {\it Phys. Lett.} {\bf B 357} 699 (1995);\\
OPAL Collaboration, Alexander G et al {\it Phys. Lett.} {\bf B 364} 93 (1995);\\
OPAL Collaboration, Akers R et al {\it Z. Phys.} {\bf C 67} 27 (1995);\\
SLD Collaboration, Abe K et al {\it Phys. Rev. Lett.} {\bf 84} 4300 (2000)
\bibitem{dyak}
Dremin I M and Yakovlev V I {\it Proc.  17 Int. Symp. on Multiparticle Dynamics}
Ed. M. Markytian, Austria, 1986 (World Scientific, Singapore, 1987) p 849
\bibitem{opa2}
OPAL Collaboration, Ackerstaff K et al {\it Eur. Phys. J.} {\bf C 7} 369 (1999)
\bibitem{nomo}
DELPHI Collaboration, Nomokonov V {\it Proc. Int. Europhysics Conf. HEP99},
Tampere, Finland, July 1999; hep-ex/9910059
\bibitem{dr79}
Dremin I M {\it Pis'ma v ZhETF} {\bf 30} 154 (1979), {\it JETP Lett.} {\bf 30}
140 (1979); {\it YaF} {\bf 33} 1357 (1981), {\it Sov. J. Nucl. Phys.} {\bf 33}
726 (1981)
\bibitem{dikk}
Dremin I M, Ivanov O V, Kalinin S A et al {\it Phys. Lett.} {\bf B 499} 97 (2001)
\bibitem{47}
Chudakov A E {\it Izv. AN SSSR, ser fiz} {\bf 19} 650 (1955)
\bibitem{115}
DELPHI Collaboration, Abreu P et al {\it Z. Phys.} {\bf C 56} 63 (1992)
\bibitem{188}
Gary J W {\it Phys. Rev.} {\bf D 61} 114007 (2000)
\bibitem{agsj}
Andersson ', Gustafson G and Sj\"{o}strand T {\it Phys. Lett.} {\bf B 94} 211 (1980) 
\bibitem{64}
Azimov Ya I, Dokshitzer Yu L, Khoze V A and Troyan S I {\it Phys. Lett.} {\bf B 165}
147 (1985); {\it Ÿ"} {\bf 43} 149 (1986); {\it Sov. J. Nucl. Phys.} {\bf 43} 95 (1986)
\bibitem{del2}
DELPHI Collaboration, Abreu P et al {\it Z. Phys.} {\bf C 70} 179 (1996)
\bibitem{155}
Dokshitzer Yu L, Khoze V A and Troyan S I {\it YaF} {\bf 46} 1220 (1987);
{\it Sov. J. Nucl. Phys.} {\bf 46} 712 (1987)
\bibitem{171}
DELPHI Collaboration, CERN-OPEN-2000-134
\bibitem{bene}
Beneke M {\it Phys Rep} {\bf 317} 1 (1999)
\bibitem{yura}
Dokshitzer Yu L Talk at {\it IPPP Workshop on Multiparticle Production in
QCD Jets}, Durham, England, Dec. 2001
\bibitem{hama}
Hamacher K Talk at {\it IPPP Workshop on Multiparticle Production in
QCD Jets}, Durham, England, Dec. 2001
\bibitem{162}
Khoze V A and Stirling W J {\it Z. Phys.} {\bf C 76} 59 (1997)
\bibitem{182}
D0 Collaboration, Abbott B et al {\it Phys. Lett.} {\bf B 414} 419 (1997) 
\bibitem{180}
Butterworth J M, Khoze V A and Ochs W {\it J. Phys.} {\bf G 25} 1457 (1999)
\bibitem{66}
Bialas A and Peschanski R {\it Nucl. Phys.} {\bf B 273} 703 (1986)
\bibitem{drje}
Dremin I M {\it Pis'ma v ZhETF} {\bf 45} 505 (1987); {\it JETP Lett.}
{\bf 45} 643 (1987) 
\bibitem{14}
DeWolf E A, Dremin I M and Kittel W {\it Phys. Rep.} {\bf 270} 1 (1996)
\bibitem{38}
Dokshitzer Yu L and Dremin I M {\it Nucl. Phys.} {\bf B 402} 139 (1993) 
\bibitem{68}
Ochs W and Wosiek J {\it Phys. Lett.} {\bf B 289} 159 (1992); {\bf B 304} 144 (1993)
\bibitem{70}
Brax Ph, Meunier J L and Peschanski R {\it Z. Phys.} {\bf C 62} 649 (1994)
\bibitem{opal}
OPAL Collaboration, Abbiendi G et al {\it Eur. Phys. J.} {\bf C 11} 239 (1999)
\bibitem{1121}
Andersson B, Dahlquist P and Gustafson G {\it Phys. Lett.} {\bf B 214} 604 (1988)
\bibitem{1122}
Dahlquist P, Andersson B and Gustafson G {\it Nucl. Phys.} {\bf B 238} 76 (1989)
\bibitem{dlne}
Dremin I M, Lam C S and Nechitailo V A {\it Phys. Rev.} {\bf D 61} 074020 (2000)
\bibitem{173}
Lupia S, Ochs W and Wosiek J {\it Nucl. Phys.} {\bf B 540} 405 (1999)
\bibitem{151}
OPAL Collaboration, Ackerstaff K et al {\it Eur. Phys. J.} {\bf C 1} 479 (1998)
\bibitem{111}
Ochs W and Wosiek J {\it Phys. Lett.} {\bf B 304} 144 (1993)
\bibitem{117}
DELPHI Collaboration, Abreu P et al {\it Phys. Lett.} {\bf B 440} 203 (1998)
\bibitem{118}
ZEUS Collaboration,  Breitweg et al {\it Eur. Phys. J.} {\bf C 12} 53 (2000)
\bibitem{119}
Dokshitzer Yu L, Khoze V A, Marchesini G and Webber B R {\it Phys. Lett.}
{\bf B 245} 243 (1990)
\bibitem{120}
Ochs W and Wosiek J {\it Z. Phys.} {\bf C 72} 263 (1996)
\bibitem{201}
Sterman G and Weinberg S {\it Phys. Rev. Lett.} {\bf 39} 1436 (1977)
\bibitem{211}
Stirling W J {\it J. Phys.} {\bf G 17} 1567 (1991)
\bibitem{221}
Sj\"{o}strand T {\it Comp. Phys. Comm.} {\bf 28} 229 (1983)
\bibitem{261}
JADE Collaboration, Bartel W et al {\it Z. Phys.} {\bf C 33} 23 (1986);
Bethke S et al {\it Phys. Lett.} {\bf B 213} 235 (1988)
\bibitem{271}
Dokshitzer Yu L, Leder G D, Moretti S and Webber B R {\it JHEP} {\bf 9708}
1 (1997)
\bibitem{l3}
L3 Collaboration, Adriani O et al {\it Phys. Rep.} {\bf 236} 1 (1993)
\bibitem{opa1}
OPAL Collaboration, Akers R et al {\it Z. Phys.} {\bf C 63} 363 (1994) 
\bibitem{seym}
Seymour M H {\it Phys. Lett.} {\bf B 378} 279 (1996)
\bibitem{alep}
ALEPH Collaboration, Busculic D et al {\it Phys. Lett.} {\bf B 346} 389 (1995) 
\bibitem{085}
H1 Collaboration, Adloff C et al {\it Nucl. Phys.} {\bf B 504} 3 (1997)
\bibitem{dine}
Dremin I M, Ivanov O V and Nechitailo V A {\it UFN}
{\bf 171} 465 (2001); {\it Phys.-Uspekhi} {\bf 44} (5) (2001).
\bibitem{glad}
Gladysz-Dziadus E {\it Elem. Part. Atom. Nucl.} (2002) (to be published)

\end{thebibliography}
\end{document}